\documentclass[envcountsame,runningheads,orivec]{llncs}
\usepackage[mathlines]{lineno}

\usepackage{amsmath,amssymb}
\usepackage[bold,basic]{complexity}
\newclass{\PiTwoP}{\mathbf{\Pi}_2^{\P}}
\newclass{\SigmaTwoP}{\mathbf{\Sigma}_2^{\P}}
\usepackage{mathtools}
\usepackage[hidelinks]{hyperref}
\usepackage[capitalise,noabbrev,nameinlink]{cleveref}
\usepackage{xspace}
\usepackage{braket}
\usepackage{todonotes}
\usepackage{etoolbox}
\usepackage{tikz}
\usepackage{wrapfig}
\usepackage{graphicx}
\usepackage{tikz}
\usepackage{xparse}
\usetikzlibrary{shapes.misc, calc, arrows.meta, decorations.markings, fit, backgrounds}
\usepackage{booktabs}
\usepackage{caption}
\usepackage{float}

\patchcmd{\qed}{\endgraf}{\par}{}{%
  \PackageError{main}{Unable to patch LNCS proof line numbering}
    {Check the definition of \string\qed.}%
}

\makeatletter
\@ifpackagelater{lineno}{2022/07/29}{}{%
  \newcommand*\patchMathForLineno[2]{%
    \csletcs{LNsaved#2}{#2}%
    \csletcs{LNsavedend#2}{end#2}%
    \renewenvironment{#2}
      {#1\csuse{LNsaved#2}}
      {\csuse{LNsavedend#2}\endlinenomath}%
  }
  \AtBeginDocument{%
    \let\linenomathAMS\linenomath
    \patchcmd{\linenomathAMS}{\advance\postdisplaypenalty\linenopenalty}{}{}{}%
    \forcsvlist{\patchMathForLineno{\linenomath}}{equation,equation*}%
    \forcsvlist{\patchMathForLineno{\linenomathAMS}}{align,align*,alignat,alignat*,gather,gather*,multline,multline*,flalign,flalign*}%
  }
}
\makeatother

\renewcommand{\vec}[1]{\boldsymbol{#1}}

\newcommand{\dom}{\mathrm{dom}}
\newcommand{\src}{\mathrm{top}}
\newcommand{\trg}{\mathrm{bot}}
\newcommand{\cho}{\mathrm{choice}}

\newcommand{\inc}{\operatorname{Inc}}

\newcommand{\N}{\mathbb{N}}
\newcommand{\cA}{\mathcal{A}}
\newcommand{\cB}{\mathcal{B}}
\renewcommand{\cL}{\mathcal{L}}

\renewcommand{\O}{\mathcal{O}}

\pgfkeys{
  /transitionSector/.is family, /transitionSector,
  default/.style = {
    shift = {},
    ratio = 0.4,
    bend left = left,
    bend right = right,
    angle left = 30,
    angle right = 30,
    sector dist = 0.7cm,
    symbol dist = 0.5cm,
    stem bend = left,
    stem angle = 0,
    label fill = white,
    apex factor = 1.25,
  },
  shift/.estore in = \sectorShift,
  ratio/.estore in = \sectorRatio,
  bend left/.estore in = \sectorBendL,
  bend right/.estore in = \sectorBendR,
  angle left/.estore in = \sectorAngleL,
  angle right/.estore in = \sectorAngleR,
  sector dist/.estore in = \sectorDist,
  symbol dist/.estore in = \symbolDist,
  stem bend/.estore in = \sectorStemBend,
  stem angle/.estore in = \sectorStemAngle,
  label fill/.estore in = \sectorLabelFill,
  apex factor/.estore in = \sectorApex,
}

\NewDocumentCommand{\transitionSector}{O{} m m m m m}{%
    \pgfkeys{/transitionSector, default, #1}%
    \if\relax\detokenize\expandafter{\sectorShift}\relax
        \coordinate (S_fork) at ($(#2)!\sectorRatio!($(#3)!0.5!(#4)$)$);
    \else
        \coordinate (S_fork) at ($(#2)+(\sectorShift)$);
    \fi
    \path[postaction={decorate, decoration={markings,
            mark=at position 0.2*\sectorDist with {\coordinate (S_L1);},
            mark=at position 0.4*\sectorDist with {\coordinate (S_L2);},
            mark=at position 0.6*\sectorDist with {\coordinate (S_L3);},
            mark=at position 0.8*\sectorDist with {\coordinate (S_L4);},
            mark=at position \sectorDist with {\coordinate (S_arcL);},
            mark=at position \symbolDist with {\coordinate (S_ptL);}}}]
          (S_fork) to[bend \sectorBendL=\sectorAngleL] (#3);
    \path[postaction={decorate, decoration={markings,
            mark=at position 0.2*\sectorDist with {\coordinate (S_R1);},
            mark=at position 0.4*\sectorDist with {\coordinate (S_R2);},
            mark=at position 0.6*\sectorDist with {\coordinate (S_R3);},
            mark=at position 0.8*\sectorDist with {\coordinate (S_R4);},
            mark=at position \sectorDist with {\coordinate (S_arcR);},
            mark=at position \symbolDist with {\coordinate (S_ptR);}}}]
          (S_fork) to[bend \sectorBendR=\sectorAngleR] (#4);
    \coordinate (S_midArc) at ($(S_arcL)!0.5!(S_arcR)$);
    \coordinate (S_apex) at ($(S_fork)!\sectorApex!(S_midArc)$);
    \fill[teal!35, fill opacity=0.3] (S_fork) -- (S_L1) -- (S_L2) -- (S_L3) -- (S_L4) -- (S_arcL)
                  .. controls ($(S_arcL)!0.5!(S_apex)$) and ($(S_arcR)!0.5!(S_apex)$) .. (S_arcR)
                  -- (S_R4) -- (S_R3) -- (S_R2) -- (S_R1) -- cycle;
    \draw[thick] (#2) to[bend \sectorStemBend=\sectorStemAngle] node[pos=0.5, font=\tiny, fill=\sectorLabelFill, inner sep=1pt] {#5} (S_fork);
    \draw[thick, dashed, -{Stealth[scale=0.9]}] (S_fork) to[bend \sectorBendL=\sectorAngleL] (#3);
    \draw[thick, -{Stealth[scale=0.9]}] (S_fork) to[bend \sectorBendR=\sectorAngleR] (#4);
    \node[font=\tiny] at ($(S_ptL)!0.5!(S_ptR)$) {#6};
}

\title{On the Complexity of Level-Synchronized Tree Automata for Quantum Circuit Verification}
\titlerunning{On the Complexity of LSTAs for Quantum Circuit Verification}
\author{Michael Blondin\inst{1} \and Samy Chady Khoumsi-Kasmi\inst{1}
\and Guillermo A. P\'erez\inst{2} \and Marnix Suilen\inst{2}}
\authorrunning{M.~Blondin et al.}
\institute{Université de Sherbrooke, Canada \and University of Antwerp, Belgium}

\begin{document}

\maketitle
\begin{abstract}
    Reliable quantum computation requires methods for verifying quantum circuits.
    Level-synchronized tree automata (LSTAs) were recently introduced to represent sets of quantum states for this purpose. We study LSTAs along two axes relevant to verification: the complexity of their decision problems and the size of their representations under circuit operations.
    We prove that acyclic LSTAs have \NP-complete nonemptiness and \PiTwoP-complete language inclusion, and that general language inclusion is \EXPSPACE-complete.
    We further prove inherent quadratic representation growth under individual quantum gates in a fixed qubit order, and give an adjacent-qubit reordering construction that preserves level synchronization.
    These results clarify the scope of applicability of the automata-theoretic toolbox for quantum circuit verification.
\end{abstract}

\section{Introduction}
\label{sec:introduction}

\begin{wrapfigure}{r}{0.42\textwidth} 
\vspace{-15pt}
\centering
\begin{tikzpicture}[
    scale=1, transform shape, 
    level distance=1.2cm,
    level 1/.style={sibling distance=2.5cm},
    level 2/.style={sibling distance=1.2cm},
    left branch/.style={edge from parent/.style={dashed, draw}},
    right branch/.style={edge from parent/.style={solid, draw}}
]

\node {$v_1$}
    child[left branch] {
        node {$v_2$}
        child[left branch] { node {$\frac{1}{2}$} }
        child[right branch] { node {$\frac{1}{2}$} }
    }
    child[right branch] {
        node {$v_2$}
        child[left branch] { node {$\frac{1}{2}$} }
        child[right branch] { node {$\frac{1}{2}$} }
    };

\end{tikzpicture}
\caption{The quantum state $\ket{++} = \frac{1}{2}(\ket{00} + \ket{01} + \ket{10} + \ket{11})$ as a perfect binary tree. Dashed and solid edges encode $0$ and $1$ values.}
\vspace{-15pt}
\label{fig:plusplus}
\end{wrapfigure}
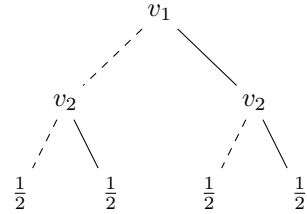

Verifying  
a quantum circuit requires reasoning about the states it can produce from its allowed inputs.
An $n$-qubit state has $2^n$ \emph{amplitudes}, and a specification may describe many such states at once. \Cref{fig:plusplus} illustrates how one such quantum state can be represented as a perfect binary tree.
\emph{Level-synchronized tree automata (LSTAs)} represent sets of quantum states by sharing subtrees and coordinating choices across each tree level~\cite{DBLP:journals/pacmpl/AbdullaCCHLLLT25}.
Synchronization encodes dependencies between distant branches that ordinary tree automata can represent only at an exponential cost (see, e.g., \cite[Ex.~3]{DBLP:journals/pacmpl/AbdullaCCHLLLT25}).
These representations underlie \textsc{AutoQ}, an LSTA-based tool for quantum-circuit verification, which supports single-qubit and controlled-gate transformations with at most quadratic representation growth~\cite{DBLP:journals/pacmpl/AbdullaCCHLLLT25}.

Computing a specified output amplitude of a circuit, known as exact strong simulation, is a closely related problem. For example, de Colnet et al.~\cite{deColnetEtAl2026SumsOfPowers} express amplitudes as weighted sums over Boolean path variables and exploit the structure of the resulting interaction graph.
Instead, LSTA-based verification symbolically simulates the circuit:
starting from an LSTA representing the allowed inputs, it transforms
the automaton for each successive gate.
It then checks whether the resulting language is included in that
of an LSTA representing the postcondition.
In this setting, representation size determines the cost of each gate transformation and of the final inclusion check.
Tsai et al.~\cite{TsaiCL26} use qubit reordering to compile pre- and postconditions into LSTAs whose size is linear in the number of qubits for fixed specification parameters.
The resulting smaller automata also reduce verification time.
This motivates our study of representation growth under gates, the cost of reordering, and the complexity of inclusion.

The automata used for quantum verification have also developed beyond the original LSTA model.
\textsc{AutoQ}~2.0 uses symbolic amplitudes to handle measurement and classical control flow with supplied loop invariants~\cite{ChenEtAl2025AutoQ2}.
More recently, \emph{synchronized weighted tree automata (SWTAs)} combine level synchronization with weighted transitions and associate quantum states with sequences of \emph{colors} (called \emph{choices} here)~\cite{AbdullaEtAl2026Parameterized}.
Their functional-inclusion test compares the outputs associated with the same color sequence.

\paragraph*{Contributions.}
Language inclusion for LSTAs is known to be \PSPACE-hard and in \EXPSPACE~\cite{DBLP:journals/pacmpl/AbdullaCCHLLLT25}.
For circuit verification with a fixed number of qubits, all state trees have the same height, so acyclic LSTAs suffice to represent the relevant sets of quantum states.
We settle the complexity of nonemptiness and inclusion for acyclic LSTAs and of inclusion for general LSTAs.
We also establish lower bounds on representation growth under quantum gates, and give an efficient construction for swapping adjacent qubits in the qubit order.
Specifically:
\begin{enumerate}
\item For acyclic LSTAs, nonemptiness is \NP-complete (\cref{thm:acyclic-nonemptiness}) and inclusion is \PiTwoP-complete (\cref{thm:acyclic-inclusion}). By contrast, nonemptiness for general LSTAs was already known to be \PSPACE-complete~\cite{DBLP:journals/pacmpl/AbdullaCCHLLLT25}.
\item Language inclusion for LSTAs is \EXPSPACE-complete (\cref{thm:lsta-inclusion-expspace}).
Hardness already holds with one initial state and singleton choice sets.
\item In a fixed qubit order, a Hadamard gate or a controlled-NOT gate whose control follows its target can force every LSTA accepting the output tree to have quadratically many states relative to the input representation size (\cref{prop:hadamard-growth,prop:backward-cx-growth}).
The latter extends to controlled-$U$ gates, for every fixed non-diagonal one-qubit unitary $U$ (\cref{cor:backward-unitary-growth}).
\item We show that swapping two adjacent qubits in the qubit order of a layered LSTA is efficient, as with adjacent-variable swaps in binary decision diagrams (BDDs)~\cite{Rudell1993Ordering}.
The construction takes linear time in the automaton size for a fixed number of available choices (\cref{prop:adjacent-reorder}).
\end{enumerate}

We work throughout with perfect binary trees, as required to represent fixed-length quantum state vectors.
\cref{subsec:complementation} also gives a self-contained nonclosure proof for this tree domain.
Classical variable-order sensitivity transfers from BDDs to LSTAs through residual counting~\cite{Bryant1986Graphs,KrauseSavickyWegener1999}, as explained in \cref{subsec:variable-order}.
Together with the gate lower bounds, it motivates careful selection of qubit orders and transformation sequences.
In the conclusion (\cref{sec:conclusion}), we identify blocks that could be applied in linear time and discuss how a verifier may combine them with dynamic reordering and evaluation from either end of a circuit. Due to space constraints, some proofs and examples are deferred to the appendix.

\section{Preliminaries}

Given a predicate $P$, we write $[P]=1$ if $P$ holds and $[P]=0$ otherwise.
Recall that $\PiTwoP$ is the class of problems whose complement belongs to $\SigmaTwoP=\NP^\NP$. A canonical $\PiTwoP$-complete problem asks whether a quantified Boolean formula $\forall X\, \exists Y\, \varphi(X,Y)$ is true, where $\varphi$ is quantifier-free and in 3-CNF.

\subsection{Quantum states}
An $n$-qubit (pure) quantum state can be understood as a function $f \colon \{0,1\}^n \to \mathbb{C}$ from bitstring \emph{addresses} to complex \emph{amplitudes}, with $\sum_{\vec{b} \in \{0,1\}^n} |f(\vec{b})|^2 = 1$.
That is, the sum of the squared moduli of its amplitudes is $1$. Equivalently, using bra-ket notation, a quantum state is a normalized vector $\ket{\psi} = \sum_{\vec{b} \in \{0,1\}^n} \alpha_{\vec{b}} \ket{\vec{b}}$, where $\alpha_{\vec{b}}=f(\vec{b})$ and $\sum_{\vec{b} \in \{0,1\}^n}|\alpha_{\vec{b}}|^2=1$.

Fixing a qubit order $\pi=(v_1,\dots,v_n)$ identifies this vector with a tree\footnote{The special case where amplitudes belong to $\{0, 1\}$ yields a binary decision tree.} of height $n$: each internal node at depth $i$ is labeled $v_{i+1}$, its two children correspond to setting that bit to $0$ or $1$, and the leaf at \emph{address} $\vec{b}$ is labeled $\alpha_{\vec{b}}$. For example, the state $\frac{1}{2}\ket{00}+\frac{1}{2}\ket{01}+\frac{1}{2}\ket{10}+\frac{1}{2}\ket{11}$ has four leaves labeled $1/2$ (see \Cref{fig:plusplus}).

\subsection{Perfect binary trees}

Let $\Sigma$ be a finite \emph{ranked alphabet}, i.e., a finite alphabet with a ranking function $\mathit{rk} \colon\Sigma\to\{0,2\}$.
Let $\Sigma_r=\{\sigma\in\Sigma : \mathit{rk}(\sigma)=r\}$ for $r\in\{0,2\}$. For an address $v \in \{0,1\}^*$ and $i \in \{0,1\}$, write $vi$ for the address obtained by appending $i$ to $v$.

\begin{definition}%
    A \emph{binary tree over $\Sigma$} is a map $T\colon D\to\Sigma$, where $D\subseteq\{0,1\}^*$ is finite, nonempty, and prefix-closed: for all $u,v\in\{0,1\}^*$, $uv\in D$ implies $u\in D$.
    For every $v\in D$, its children are those prescribed by its label:
    \[
    \{i\in\{0,1\}:v i\in D\}=
    \begin{cases}
        \emptyset,& \text{if }T(v)\in\Sigma_0,\\
        \{0,1\},& \text{if }T(v)\in\Sigma_2.
    \end{cases}
    \]
\end{definition}

The \emph{depth} of a node $v\in D$ is its length $|v|$.
The height of the tree is $h(T)=\max\{|v|:v\in D\}$, and its $n$th \emph{level} is $D_n=\{v\in D:|v|=n\}$.
A tree is \emph{perfect} if all its leaves have the same depth, equivalently, if $D=\{0,1\}^{\leq h(T)}$.
Write $\operatorname{Perf}(\Sigma)$ for the set of all perfect binary trees over $\Sigma$.
Height~$0$ is allowed: such a tree is a single rank-$0$ root $\sigma$, with $T$ mapping the empty word $\varepsilon$ to $\sigma$.

For $x$ in the domain $\dom(T)$ of a tree $T$, the subtree $T^x$ is defined by
\[
\dom(T^x)=\{u\in\{0,1\}^*:xu\in\dom(T)\},
\qquad T^x(u)=T(xu).
\]
For $u,v\in\dom(T)$, we write $u\equiv_T v$ whenever $T^u=T^v$
as partial functions.

\subsection{Level-synchronized tree automata}
We use the LSTA model of Abdulla et al.~\cite{DBLP:journals/pacmpl/AbdullaCCHLLLT25} on finite perfect binary trees.

\begin{definition}%
A \emph{level-synchronized tree automaton (LSTA)} is a tuple $\cA=(Q,I,\Sigma,\Delta)$ consisting of a finite state set $Q$, initial states $I\subseteq Q$, a finite ranked alphabet $\Sigma$, and a finite set of transitions $\Delta$ of the forms
\[
q\xrightarrow{C}\sigma(q_0,q_1) \text{ with } \sigma\in\Sigma_2, \text{ or } q\xrightarrow{C}\sigma \text{ with } \sigma\in\Sigma_0,
\]
where $q,q_0,q_1 \in Q$ and $C \subseteq \N$ is finite. These are called \emph{internal} and \emph{leaf} transitions, respectively.
The transitions must be \emph{choice-deterministic}: distinct transitions with the same source have disjoint choice sets.
\end{definition}

Let $\delta \in \Delta$. For either form, let $\src(\delta) = q$ and $\cho(\delta)=C$. For internal transitions, let $\trg(\delta)=\{q_0,q_1\}$; for leaf transitions, let $\trg(\delta)=\emptyset$. For an LSTA $\cA$, we write
$\cA=(Q_{\cA},I_{\cA},\Sigma_{\cA},\Delta_{\cA})$
to identify its components.

A \emph{run} of $\cA$ on a perfect binary tree $T\colon D\to\Sigma$ is a map $\rho\colon D\to\Delta$ labeling its nodes by transitions, subject to the following conditions:
\begin{itemize}
    \item At each leaf $v$, the transition $\rho(v)$ has the form $q\xrightarrow{C}T(v)$.
    \item At each internal node $v$, the transition $\rho(v)$ has the form $q\xrightarrow{C}T(v)(q_0,q_1)$, with $q_i=\src(\rho(vi))$ for $i\in\{0,1\}$.
\end{itemize}
The key feature of LSTAs is that all nodes at the same depth must agree on a common choice, even though they may use different transitions.
Formally, we say that a run $\rho$ on $T$ is \emph{accepting} if
\[
\src(\rho(\varepsilon))\in I
\quad\text{and}\quad
\forall n\in\{0,\ldots,h(T)\}:
\bigcap_{v\in D_n}\cho(\rho(v))\ne\emptyset.
\]

The \emph{language} of an LSTA $\cA$ is
$
\cL(\cA)=\{T\in\operatorname{Perf}(\Sigma):
\cA$ has an accepting run on $T\}$.
Fixing an initial state and one common choice per level determines at most one run and its labeled tree.
A tree may nevertheless have several accepting runs.
We call $\cA$ a \emph{singleton LSTA} if $\cL(\cA)$ contains exactly one tree.

The \emph{state-dependency graph} of an LSTA $\cA = (Q,I,\Sigma,\Delta)$ is $(Q,E_{\cA})$, where
\[
E_{\cA}=\{(q,r):\exists\delta\in\Delta \text{ s.t.\ }
\src(\delta)=q \text{ and } r\in\trg(\delta)\}.
\]
The LSTA is \emph{acyclic} if this graph is acyclic.
It is \emph{layered} if $Q$ can be partitioned into levels $Q_0,\ldots,Q_h$ with $I\subseteq Q_0$ such that, for all $i<h$, transitions from $Q_i$ are internal and have children in $Q_{i+1}$. States in $Q_h$ have only leaf transitions.

For complexity statements, size means the total length of the explicit encoding, including the transitions and every element of their choice sets.
Choice identifiers may be consistently renamed to a polynomially bounded range.
The \emph{choice universe} $K(\cA)$ contains all choices occurring on transitions, and a \emph{choice incidence} is a transition paired with one of its choices:
\[
K(\cA)=\bigcup_{\delta\in\Delta_{\cA}}\cho(\delta),\qquad
\inc(\cA)=\{(\delta,c):\delta\in\Delta_{\cA},\ c\in\cho(\delta)\}.
\]
The \emph{size of $\mathcal{A}$} is defined as
$
|\cA|=|Q_{\cA}|+|\Delta_{\cA}|+|\inc(\cA)|
$.

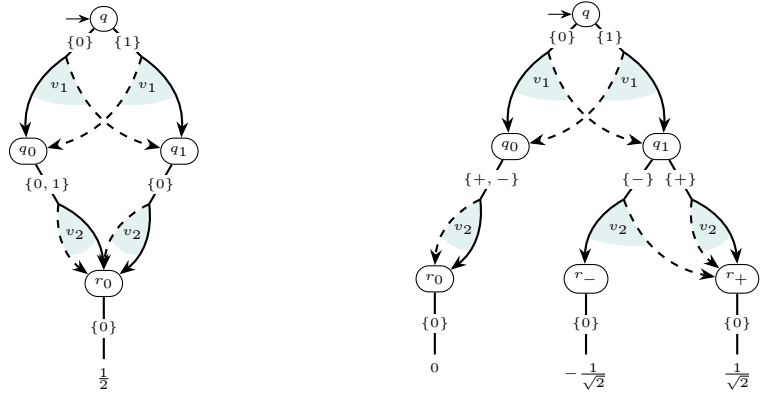
\begin{figure}
  \centering
  \begin{minipage}[b]{0.4\textwidth}
    \centering
    \begin{tikzpicture}[yscale=-1]
        \node[rounded rectangle, draw=black, font=\tiny] (q)  at (1, 0) {$q$};
        \node[rounded rectangle, draw=black, font=\tiny] (10) at (0, 1.75) {$q_0$};
        \node[rounded rectangle, draw=black, font=\tiny] (q1) at (2, 1.75) {$q_1$};
        \node[rounded rectangle, draw=black, font=\tiny] (r0) at (1, 3.5) {$r_0$};

        \draw[-{Stealth[scale=0.9]}] ($(q) + (-0.5cm, 0)$) -- (q);
        
        \transitionSector[shift={-0.5cm, 0.5cm}, bend right=left]{q}{q1}{10}{$\{0\}$}{$v_1$}
        \transitionSector[shift={0.5cm, 0.5cm}, bend left=right]{q}{10}{q1}{$\{1\}$}{$v_1$}
        \transitionSector{q1}{r0}{r0}{$\{0\}$}{$v_2$}
        \transitionSector{10}{r0}{r0}{$\{0, 1\}$}{$v_2$}
        
        \draw[thick] (r0) -- ++(0,1cm) node[midway, font=\tiny, fill=white, inner sep=1pt] {$\{0\}$} node[below, font=\tiny] {$\frac{1}{2}$};
    \end{tikzpicture}
  \end{minipage}%
  \hfill
  \begin{minipage}[b]{0.55\textwidth}
    \centering
        \begin{tikzpicture}[
            yscale=-1,
            state/.style={rounded rectangle, draw=black, font=\tiny},
            ]
            \node[state] (q)  at (2, 0) {$q$};
            \node[state] (q0) at (1, 1.75) {$q_0$};
            \node[state] (q1) at (3, 1.75) {$q_1$};
            \node[state] (r0) at (0, 3.5) {$r_0$};
            \node[state] (r1) at (2, 3.5) {$r_-$};
            \node[state] (r2) at (4, 3.5) {$r_+$};

            \draw[-{Stealth[scale=0.9]}] ($(q) + (-0.5cm, 0)$) -- (q);
            
            \transitionSector[shift={-0.5cm, 0.5cm}, bend right=left]{q}{q1}{q0}{$\{0\}$}{$v_1$}
            \transitionSector[shift={0.5cm, 0.5cm}, bend left=right]{q}{q0}{q1}{$\{1\}$}{$v_1$}
            \transitionSector{q0}{r0}{r0}{$\{+, {-}\}$}{$v_2$}
            \transitionSector[shift={-0.5cm, 0.7cm}, bend right=left]{q1}{r2}{r1}{$\{-\}$}{$v_2$}
            \transitionSector{q1}{r2}{r2}{$\{+\}$}{$v_2$}
            
            \draw[thick] (r2) -- ++(0,1cm) node[midway, font=\tiny, fill=white, inner sep=1pt] {$\{0\}$} node[below, font=\tiny] {$\frac{1}{\sqrt{2}}$};
            \draw[thick] (r1) -- ++(0,1cm) node[midway, font=\tiny, fill=white, inner sep=1pt] {$\{0\}$} node[below, font=\tiny] {$-\frac{1}{\sqrt{2}}$};
            \draw[thick] (r0) -- ++(0,1cm) node[midway, font=\tiny, fill=white, inner sep=1pt] {$\{0\}$} node[below, font=\tiny] {$0$};
        
        \end{tikzpicture}
  \end{minipage}
  \caption{Two examples of LSTAs. \emph{Left:} A singleton LSTA $\cA_1$ recognizing $\lvert ++\rangle$. \emph{Right:} An LSTA $\cA_2$ recognizing $\{\lvert 0+\rangle, \lvert 0-\rangle, \lvert 1+\rangle, \lvert 1-\rangle\}$.}
  \label{fig:2LSTAs}
\end{figure}

\begin{example}
In \Cref{fig:2LSTAs}, the LSTA $\mathcal A_1$ is nonminimal and admits different choices, but accepts only the tree $T$ of \Cref{fig:plusplus}.
Choosing $0$ at each level gives the accepting run $\rho\colon\operatorname{dom}(T)\to\Delta$ defined by
\begin{align*}
  \rho(\varepsilon) &= \phantom{{}_1}q \xrightarrow{\{0\}} v_1(q_1, q_0), \\
  \rho(0) &= q_1 \xrightarrow{\{0\}} v_2(r_0, r_0), \quad 
  \rho(1) = q_0 \xrightarrow{\{0, 1\}} v_2(r_0, r_0), \\
  \rho(00) &= \rho(01) = \rho(10) = \rho(11) = r_0 \xrightarrow{\{0\}} \tfrac{1}{2}.
\end{align*}

The LSTA $\mathcal A_2$ accepts the tree representation of $\lvert 1-\rangle = \frac{1}{\sqrt{2}}(\lvert 10\rangle-\lvert11\rangle)$ using choices\footnote{For readability, we use arbitrary symbols for choices but these can be replaced by natural-number indices.} $1$, $-$, and $0$, with the following accepting run:
\begin{align*}
  \rho(\varepsilon) &= \phantom{{}_0}q \xrightarrow{\mathmakebox[23pt][c]{\{1\}}} v_1(q_0, q_1), \\
  \rho(0) &= q_0 \xrightarrow{\mathmakebox[23pt][c]{\{+, -\}}} v_2(r_0, r_0), \quad 
  \rho(1) = q_1 \xrightarrow{\{-\}} v_2(r_+, r_-), \\
  \rho(00) &= \rho(01) = r_0 \xrightarrow{\{0\}} 0, \quad 
  \rho(10) = r_+ \xrightarrow{\{0\}} \tfrac{1}{\sqrt{2}}, \quad 
  \rho(11) = r_- \xrightarrow{\{0\}} -\tfrac{1}{\sqrt{2}}.\tag*{\qed}
\end{align*}
\end{example}

\subsection{Complementation}
\label{subsec:complementation}

\begin{theorem}
\label{thm:nonclosure-complement}
    LSTA languages are not closed under complement relative to perfect binary trees.
\end{theorem}

Abdulla et al.~\cite[Thm.~12]{DBLP:journals/pacmpl/AbdullaCCHLLLT25} proved nonclosure under complement over unrestricted finite binary trees. Since we work with perfect binary trees, this does not establish nonclosure for our setting. 
We use the following property of LSTAs. It follows directly from the assumption of choice determinism for LSTAs (see also the proof of \cite[Thm.~14]{DBLP:journals/pacmpl/AbdullaCCHLLLT25}).%

\begin{lemma}
\label{lem:equal-state-subtrees}
Let $\rho$ be an accepting run on a perfect binary tree $T$.
If $u$ and $v$ have the same depth and $\src(\rho(u))=\src(\rho(v))$, then $T^u=T^v$.
\end{lemma}

\noindent
We exhibit an LSTA whose complement, restricted to perfect trees, contains trees with arbitrarily many distinct subtrees at one depth. By \cref{lem:equal-state-subtrees}, no LSTA can recognize that complement.

\begin{proof}[of \cref{thm:nonclosure-complement}]
Let $\Sigma_2=\{x\}$ and $\Sigma_0=\{a,b\}$.
Consider the LSTA $\cA$ with sole (initial) state $q$, and transitions $q\xrightarrow{\{1\}}x(q,q)$ and $q\xrightarrow{\{2\}}a$.
It accepts exactly the perfect binary trees whose leaves are all labeled $a$, including the height-$0$ tree $a$.
Suppose that an LSTA $\cB$ recognizes its complement relative to $\operatorname{Perf}(\Sigma)$, and let $m=|Q_{\cB}|$.
Since $\cB$ must accept the height-$0$ tree $b$, we have $m\geq1$.
Consider the perfect binary tree $T$ of height $2m$, with all internal nodes labeled $x$, and such that
\[
T(uv)=\begin{cases}
b & \text{if } u=v,\\
a & \text{otherwise},
\end{cases}\quad
\text{for } u,v\in\{0,1\}^m.
\]
It has a $b$-labeled leaf and hence must be accepted by $\cB$.
The $2^m$ subtrees rooted at depth $m$ are pairwise distinct: $T^u$ has its unique $b$-labeled leaf at relative address $u$, i.e., $T^u(v) = b$ if and only if $u = v$.
As $2^m>m$, two of these roots have the same state in any accepting run of $\cB$.
\cref{lem:equal-state-subtrees} forces their subtrees to be equal, a contradiction.\qed
\end{proof}

\section{Acyclic LSTAs: nonemptiness and language inclusion}
\label{sec:nonemptiness}
\label{sec:acyclic-inclusion}
The language-inclusion problem allows to test whether all states that a quantum circuit produces from a precondition satisfy a postcondition~\cite{DBLP:journals/pacmpl/AbdullaCCHLLLT25}.
We first study nonemptiness for acyclic LSTAs, which cover the fixed-qubit applications of that work~\cite[Sect.~6.3]{DBLP:journals/pacmpl/AbdullaCCHLLLT25}.
(Recall that for general LSTAs, it is known to be \PSPACE-complete~\cite[Thm.~13]{DBLP:journals/pacmpl/AbdullaCCHLLLT25}.)
Our construction also yields hardness of inclusion.

\subsection{Nonemptiness for acyclic LSTAs}

\begin{theorem}
\label{thm:acyclic-nonemptiness}
    Nonemptiness for acyclic LSTAs is \NP-complete.
\end{theorem}

We prove \cref{thm:acyclic-nonemptiness} by establishing membership and hardness separately.
A \emph{run certificate} specifies an initial state, a height, and one common choice per level (cf.~\cite[Thm.~14]{DBLP:journals/pacmpl/AbdullaCCHLLLT25}).
A \emph{tree DAG} $D$ is a rooted directed acyclic graph representing a tree with shared subtrees.
Each node is labeled by a symbol: a rank-$0$ node has no successors, while a rank-$2$ node stores an ordered pair of successor pointers, called its children.
The pointers may have the same target, and a node may be the target of several pointers.
Its \emph{unfolding}, written $\operatorname{unfold}(D)$, recursively copies the subgraphs reached via these pointers to form a tree.
Sharing identical subtrees can make a tree DAG exponentially smaller than its unfolding.
We say that a tree DAG is \emph{layered} if its nodes can be partitioned into levels with each pointer advancing one level.

\begin{lemma}
\label{lem:nonemptiness-np}
\label{lem:acyclic-run-certificate}
Nonemptiness for acyclic LSTAs is in \NP.
A run certificate yields, in polynomial time, a layered tree DAG whose unfolding is the accepted tree.
\end{lemma}
\begin{proof}[sketch]
    Acyclicity bounds the height of any accepting run by $h<|Q|$.
    We can thus guess a run certificate: an initial state, a height, and a common choice at each level.
    At each level, choice determinism gives at most one transition for each reachable state and the chosen common choice.
    We check that these transitions exist and have the required arity, then propagate child states at internal levels.
    If we succeed, the selected transitions form a layered tree DAG with at most $|Q|^2$ nodes whose unfolding is accepted. Conversely, any accepting run yields such a certificate. As all choices occur explicitly in the automaton, both guessing and verification take polynomial time.\qed
\end{proof}

Let $\varphi$ be a CNF formula over variables
$z_1,\ldots,z_N$ with clauses $D_1,\ldots,D_m$,
where $m,N\geq1$.
Let $\bullet,a_0,a_1$ be distinct rank-$2$
symbols, and $\#$ a rank-$0$ symbol.
Symbols $a_0$ and $a_1$ encode Boolean values.
We encode assignments as trees exposing a chosen
set of variables $V\subseteq\{1,\ldots,N\}$.
We will take $V=\emptyset$ for non\-emptiness, and will expose the universally quantified variables for inclusion.
Put $d=\lceil\log_2 m\rceil$.
For every $u\in\{0,1\}^V$, define $T_u^V$
to have $d$ routing levels labeled $\bullet$,
followed by $N$ variable levels and leaves
labeled $\#$.
The routing levels provide $2^d$ branches for checking the clauses in parallel, repeating clauses if necessary.
Each node at variable level $i$ has label
$a_{u(i)}$ if $i\in V$, and $\bullet$ otherwise.
For a full assignment $\beta\in\{0,1\}^N$,
write $\beta|_V$ for its restriction to $V$.

\begin{lemma}
\label{lem:assignment-projection}
There exists a layered LSTA $\mathcal S_{\varphi,V}$ of size $\O(mN)$ such that
\begin{equation}
\cL(\mathcal S_{\varphi,V})
=\left\{T_{\beta|_V}^{V}:\beta\in\{0,1\}^N,\ \varphi(\beta)=1\right\}.
\label{eq:assignment-projection}
\end{equation}
\end{lemma}
\begin{proof}
We repeat clauses as necessary to obtain $p=2^d<2m$
clauses $D_1,\ldots,D_p$.
We first construct a routing gadget in the LSTA:
a perfect binary tree of states of depth $d$, rooted
at the initial state, whose $p$ states at depth $d$
are $(j,0,0)$ for $1\leq j\leq p$.

Each internal routing state has a single transition
$q\xrightarrow{\{0\}}\bullet(q_0,q_1)$
to its two children.
Thus, the first $d$ levels of every accepted tree
are labeled $\bullet$.
For $d=0$, the initial state is $(1,0,0)$
and no routing transitions are needed.

After the initial routing gadget, we want states $(j,i,s)$ to record whether the first $i$ assigned variables have already satisfied $D_j$.
For $i \in \{1,\ldots,N\}$ and $s,b\in\{0,1\}$, add transitions
\[
(j,i-1,s)\xrightarrow{\{b\}}
\ell_i^V(b)\bigl((j,i,s'),(j,i,s')\bigr) \text{ with }
\ell_i^V(b)=\begin{cases}a_b& \text{if } i\in V,\\
\bullet &\text{otherwise},\end{cases}
\]
where $s'$ is $1$ if either $s=1$ or setting $z_i$ to $b$ satisfies  clause $D_j$.
Only $(j,N,1)$ has a leaf transition, with label $\#$ and choice $\{0\}$. An example of the construction with $V = \emptyset$ is illustrated in \Cref{nonempty_acyclic_figure}.

At each variable level $i$, synchronization selects
one bit $b$ for all clause components.
A level is \emph{visible} if $i\in V$, in which case
its label is $a_b$; otherwise its label is $\bullet$,
but the choice still updates every component.

Note that the construction is layered and choice-deterministic,
with $\O(pN+p)=\O(mN)$ states, transitions, and
choice incidences.

For any assignment $\beta\in\{0,1\}^N$, an induction on $i$
shows that the flag $s$ of component $j$ is $1$ after
level $i$ exactly when the first $i$ variables
satisfy $D_j$.
Thus, the choices prescribed by $\beta$ yield an accepting
run exactly when $\varphi(\beta)=1$, and the resulting
tree is $T_{\beta|_V}^V$.
Conversely, every accepting run determines such
a satisfying assignment.
This proves~\cref{eq:assignment-projection}.\qed
\end{proof}

\noindent
\begin{minipage}[t]{0.49\textwidth}
\vspace{-7pt}
\begin{example}
    Consider the LSTA of \Cref{nonempty_acyclic_figure}. 
    The first level (within a rounded box) represents the sole routing level of the gadget. 
    The subsequent second and third levels encode the truth assignments to $x_1$ and $x_2$. 
    The highlighted (blue) states illustrate an accepting run reached by the choice sequence $0, 0, 1, 0$, which sets $x_1$ to $\mathrm{false}$ and $x_2$ to $\mathrm{true}$, thereby satisfying the formula. \qed
\end{example}
Recall that in the inclusion reduction, the visible variables will be the universally quantified ones. For nonemptiness hardness, we take $V=\emptyset$.
\begin{corollary}
\label{lem:nonemptiness-hardness}
Testing nonemptiness for acyclic LSTAs is \NP-hard, even with one initial state, one rank-$2$ symbol, and one rank-$0$ symbol.
\end{corollary}
\end{minipage}\hfill
\begin{minipage}[t]{0.5\textwidth}
\vspace{-10pt}
\vspace{0pt}
\centering
\begin{tikzpicture}[
    yscale=-1,
    y=0.9cm,
    x=0.8cm,
    state/.style={rounded rectangle, draw=black, font=\tiny},
    acc/.style={fill=blue!30},
    ]
    \node[state, acc] (q) at (2.5, 0) {$q$};
    
    \node[state, acc] (000) at (0.5, 2) {$1, 0, 0$};
    \node[state, acc] (100) at (4.5, 2) {$2, 0, 0$};
    
    \node[state, acc] (010) at (0, 4) {$1, 1, 0$};
    \node[state] (011) at (2, 4) {$1, 1, 1$};
    \node[state] (110) at (4, 4) {$2, 1, 0$};
    \node[state, acc] (111) at (6, 4) {$2, 1, 1$};
    
    \node[state] (020) at (0, 6) {$1, 2, 0$};
    \node[state, acc] (021) at (2, 6) {$1, 2, 1$};
    \node[state] (120) at (4, 6) {$2, 2, 0$};
    \node[state, acc] (121) at (6, 6) {$2, 2, 1$};

    \draw[-{Stealth[scale=0.9]}] ($(q) + (-0.5cm, 0)$) -- (q);
    \transitionSector[label fill=gray!15, symbol dist = 0.8cm, apex factor=5]{q}{000}{100}{\{0\}}{$\bullet$}

    \transitionSector{000}{010}{010}{\{0\}}{$\bullet$}
    \transitionSector{000}{011}{011}{\{1\}}{$\bullet$}
    
    \transitionSector{010}{020}{020}{\{0\}}{$\bullet$}
    \transitionSector{010}{021}{021}{\{1\}}{$\bullet$}

    \transitionSector{100}{111}{111}{\{0\}}{$\bullet$}
    \transitionSector{100}{110}{110}{\{1\}}{$\bullet$}
    
    \transitionSector{110}{120}{120}{\{0\}}{$\bullet$}
    \transitionSector{110}{121}{121}{\{1\}}{$\bullet$}

    \transitionSector{011}{021}{021}{\{0, 1\}}{$\bullet$}
    \transitionSector{111}{121}{121}{\{0, 1\}}{$\bullet$}

    \draw[thick] (021) -- ++(0,1cm) node[midway, font=\tiny, fill=white, inner sep=1pt] {\{0\}} node[below, font=\tiny] {\#};
    \draw[thick] (121) -- ++(0,1cm) node[midway, font=\tiny, fill=white, inner sep=1pt] {\{0\}} node[below, font=\tiny] {\#};
    \begin{scope}[on background layer]
        \node[fit={(000 |- q.north) (100 |- 000.north)}, 
          fill=gray!15, rounded corners=10pt, inner sep=2.5pt] {};
    \end{scope}
\end{tikzpicture}
\captionof{figure}{LSTA being non empty iff the formula $(x_1 \lor x_2) \land (\neg x_1 \lor x_2)$ is satisfiable.}
\label{nonempty_acyclic_figure}
\end{minipage}
\begin{proof}
Apply \cref{lem:assignment-projection} to an
arbitrary 3-SAT instance with $V=\emptyset$,
and discard the unused symbols $\{a_0,a_1\}$.\qed
\end{proof}

\subsection{Language inclusion for acyclic LSTAs}
Making universally quantified variables visible, while leaving existentially quantified ones invisible, turns quantified satisfiability into inclusion.
The upper bound uses the polynomial run certificates established earlier.
\Cref{app:acyclic-inclusion-example} presents a worked graphical example of the reduction.

\begin{theorem}
\label{thm:acyclic-inclusion}
Inclusion for acyclic LSTAs is \PiTwoP-complete.
Hardness holds with one initial state per automaton, two choices, and a fixed ranked alphabet.
\end{theorem}
\begin{proof}[sketch]
For the upper bound, noninclusion belongs to
$\SigmaTwoP$: guess a run certificate for $\cA$
and construct its polynomial-size tree DAG $D$
using \cref{lem:acyclic-run-certificate}.
Restrict $\cB$ to the tree represented by $D$
by pairing its states with nodes of $D$ and
retaining only transitions matching the labels
and ordered children of $D$.
The resulting acyclic LSTA $\cB\!\upharpoonright_D$
has polynomial size and satisfies
\[
\cL(\cB\!\upharpoonright_D)
=\cL(\cB)\cap\{\operatorname{unfold}(D)\}.
\]
An \NP{} nonemptiness oracle can now decide
whether the guessed tree is in $\mathcal{L}(\cB)$.

For hardness, we reduce from $\forall\exists$\,3-SAT.
Given $\forall X\,\exists Y\,\varphi(X,Y)$, let us
order the variables with $X$ first and put
$V=\{1,\ldots,|X|\}$.
Let $\theta$ be a tautological CNF over the same
variables, with the same number of clauses as
$\varphi$.
We apply \cref{lem:assignment-projection} twice,
obtaining $\cA=\mathcal S_{\theta,V}$ and
$\cB=\mathcal S_{\varphi,V}$.
By \cref{eq:assignment-projection},
\[
\cL(\cA)\subseteq\cL(\cB)
\iff
\forall x\in\{0,1\}^X\;
\exists y\in\{0,1\}^Y:
\varphi(x,y).
\]
Both automata have polynomial size and satisfy
the stated restrictions.\qed
\end{proof}

\section[Language inclusion is EXPSPACE-complete]
{Language inclusion is \EXPSPACE-complete}
\label{sec:expspace-inclusion}

We now allow cycles, so accepting trees need not have bounded
height. We prove \EXPSPACE-hardness by reducing from finite-trace inclusion for \emph{labeled
1-safe Petri nets}. This matches the upper bound of Abdulla
et al.~\cite[Thm.~15]{DBLP:journals/pacmpl/AbdullaCCHLLLT25}
and closes the gap left by their \PSPACE{} lower bound.

\begin{theorem}
\label{thm:lsta-inclusion-expspace}
Language inclusion for LSTAs is \EXPSPACE-complete.
\end{theorem}

The reduction uses one component for each place of the net:
a pair of automaton states recording whether that place contains a token,
together with their outgoing transitions.
After initialization, a common choice at each internal level selects a net transition,
which all components check and simulate together.
The tree records the resulting sequence of transition labels. To make this concrete, we need some definitions.

\subsection{Labeled 1-safe Petri nets and trace inclusion}
\label{subsec:safe-net-traces}

We use Petri nets with arcs of weight one, following
\cite[Def.~2.1.1]{Jategaonkar1993Observing}.
A \emph{labeled net} is a tuple
$N=(P,\mathcal T,\operatorname{pre},\operatorname{post},M_0,\lambda)$,
where $P$ is a finite set of \emph{places} and $\mathcal T$ is a finite
set of \emph{transitions} disjoint from $P$. (We reserve $T$ for trees.)
The maps $\operatorname{pre},\operatorname{post}\colon
\mathcal T\to 2^P$ specify the input and output places.
A \emph{marking} $M\colon P\to\N$ gives the number $M(p)$ of \emph{tokens}
at each place $p$, and $M_0$ is the \emph{initial marking}.
Each transition $t$ emits the letter $\lambda(t)\in\Gamma$,
where $\Gamma$ is a finite alphabet.

At a marking $M$, a transition $t$ is \emph{enabled}
when each input place contains a token, that is,
$M(p)\geq[p\in\operatorname{pre}(t)]$ for every $p\in P$.
Firing $t$ removes one token from each input place and adds
one to each output place, giving $M\xrightarrow{t}M'$ with
\[
M'(p)=M(p)-[p\in\operatorname{pre}(t)]
             +[p\in\operatorname{post}(t)].
\]
The net is \emph{1-safe} if every marking reachable from
$M_0$ takes values in $\{0,1\}$.
Its \emph{finite trace language} is
\[
\operatorname{Tr}(N)=
\left\{\lambda(t_1)\cdots\lambda(t_k):
M_0\xrightarrow{t_1}\cdots\xrightarrow{t_k}M_k,\ k\geq0\right\}.
\]

\emph{Trace inclusion} asks whether
$\operatorname{Tr}(N_1)\subseteq\operatorname{Tr}(N_2)$
for two given such nets over a common alphabet.
This problem is
\EXPSPACE-hard by the trace-containment construction given in
Jategaonkar's proof of Theorem~6.5.3~\cite[pp.~142--143]
{Jategaonkar1993Observing}; see also
\cite{JategaonkarMeyer1996SafeNets} and more recently \cite[Prop.~4.3 and Appendix~A]{BerardEtAl2018PetriNets}.

\subsection{Reduction and proof}
\label{subsec:safe-net-reduction}

\paragraph{A simple tree encoding.}
For the input nets $N_1$ and $N_2$, put
\[
r=\max\{1,|P_1|,|P_2|\},\qquad
m=2^{\lceil\log_2 r\rceil},\qquad d=\log_2 m.
\]
Add isolated, initially empty places to each net until
both have $m$ places. These places occur in no input or
output set, so they do not change either trace language.
The two automata will use this same depth $d$ and the alphabet
\[
\Sigma_2=\Gamma\uplus\{\diamond\},\qquad
\Sigma_0=\{\bot\},\qquad
\text{where $\diamond$ and $\bot$ are fresh.}
\]

For $w=a_1\cdots a_k\in\Gamma^*$, let $E_d(w)$ be the
perfect tree of height $d+k$ with label $\diamond$ at
depths less than $d$, label $a_i$ at depth $d+i-1$
for $1\leq i\leq k$, and label $\bot$ at the leaves.
For a fixed $d$, the map $E_d$ is injective: the tree height
and the labels after the initialization prefix recover $w$.
In particular, $E_d(\varepsilon)$ has height $d$.
For a language $L\subseteq\Gamma^*$, write
$E_d(L)=\{E_d(w):w\in L\}$.

\paragraph{The automaton.}
We work with either %
net $N$ and write
its places as $P=\{p_1,\allowbreak \ldots,\allowbreak p_m\}$.
Choose pairwise distinct identifiers $c_{\mathrm{init}}$,
$c_{\mathrm{stop}}$, and $c_t$ for $t\in\mathcal T$.
The LSTA $\cA_{N,d}$ has states $q_{j,0},q_{j,1}$ for each
place $p_j$, recording its token count.

Let $\alpha_1,\ldots,\alpha_m$ be the length-$d$ binary
addresses in lexicographic order.
For $d>0$, construct a routing gadget,
as in the construction of \cref{lem:assignment-projection}: routing
edges form a perfect binary tree of depth $d$, with $m-1$
fresh internal states and state
$q_{j,M_0(p_j)}$ at address $\alpha_j$. The root is the sole initial state.

Each internal state $s$ has the single transition
\(
s\xrightarrow{\{c_{\mathrm{init}}\}}\diamond(r_0,r_1),
\)
where $r_0,r_1$ are its two children in the gadget.
Hence, the first $d$ levels of the accepted trees
have label $\diamond$, and the run states at depth $d$
record the initial marking.

For every place $p_j$, bit $b\in\{0,1\}$, and net
transition $t$, add the transitions
\begin{equation}
\begin{gathered}
q_{j,b}\xrightarrow{\{c_t\}}
\lambda(t)(q_{j,b'},q_{j,b'})\\
\text{if } b\geq[p_j\in\operatorname{pre}(t)]
\text{ and }
b'=b-[p_j\in\operatorname{pre}(t)]
      +[p_j\in\operatorname{post}(t)]\in\{0,1\}.
\end{gathered}
\label{eq:net-simulation}
\end{equation}
Both children update the bit for the same place.
For every bit state, also add
\[
q_{j,b}\xrightarrow{\{c_{\mathrm{stop}}\}}\bot.
\]

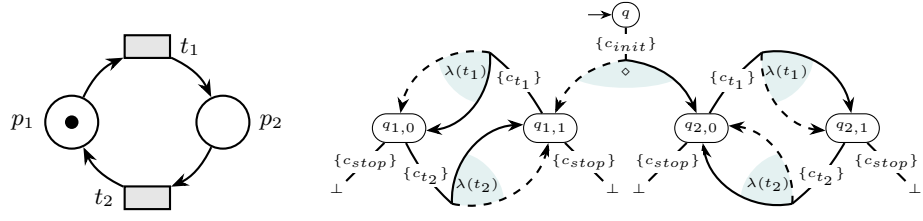
\begin{figure}
  \centering
  \begin{minipage}[b]{0.3\textwidth}
    \centering
        \begin{tikzpicture}[
            place/.style={circle, draw, thick, minimum size=7mm, inner sep=0pt},
            transition/.style={rectangle, draw, thick, fill=black!10, minimum width=6mm, minimum height=3mm},
            token/.style={circle, fill=black, inner sep=1.8pt},
            >=Stealth, thick
            ]
            \node[place, label=left:{$p_1$}] (p1) at (-1, 0) {};
            \node[place, label=right:{$p_2$}] (p2) at (1, 0) {};
            \node[token] at (p1) {};

            \node[transition, label=right:{$t_1$}] (t1) at (0, 1) {};
            \node[transition, label=left:{$t_2$}]  (t2)  at (0, -1) {};

            \draw[->] (p1) to[bend left=20] (t1);
            \draw[->] (t1) to[bend left=20] (p2);
            
            \draw[->] (p2) to[bend left=20] (t2);
            \draw[->] (t2) to[bend left=20] (p1);
        \end{tikzpicture}
  \end{minipage}%
  \hfill
  \begin{minipage}[b]{0.65\textwidth}
    \centering
        \begin{tikzpicture}[
            yscale=-1,
            state/.style={rounded rectangle, draw=black, font=\tiny},
            ]
            \node[state] (q)  at (0, 0) {$q$};
            \node[state] (p10) at (-3, 1.5) {$q_{1, 0}$};
            \node[state] (p20) at (1, 1.5) {$q_{2, 0}$};
            \node[state] (p11) at (-1, 1.5) {$q_{1, 1}$};
            \node[state] (p21) at (3, 1.5) {$q_{2, 1}$};

            \draw[-{Stealth[scale=0.9]}] ($(q) + (-0.5cm, 0)$) -- (q);
            
            \transitionSector[apex factor=2]{q}{p11}{p20}{$\{c_{init}\}$}{$\diamond$}

            \transitionSector[shift={-0.8cm, -1cm}, angle left=45, angle right=45, stem angle=10, apex factor=1.5]{p11}{p10}{p10}{\{$c_{t_1}$\}}{$\lambda(t_1)$\;}
            \transitionSector[shift={0.8cm, -1cm}, angle left=45, angle right=45, stem angle=-10, apex factor=1.5] {p20}{p21}{p21}{\{$c_{t_1}$\}}{\;$\lambda(t_1)$}

            \transitionSector[shift={0.7cm, 1cm}, angle left=45, angle right=45, stem angle=10, apex factor=1.5]{p10}{p11}{p11}{\{$c_{t_2}$\}}{\;$\lambda(t_2)$}
            \transitionSector[shift={-0.8cm, 1cm}, angle left=45, angle right=45, stem angle=-10, apex factor=1.5]{p21}{p20}{p20}{\{$c_{t_2}$\}}{$\lambda(t_2)$\;}

            \draw[thick] (p10) -- ++(-0.7cm, 0.7cm) 
            node[midway, font=\tiny, fill=white, inner sep=1pt] {$\{c_{stop}\}$} 
            node[anchor=north east, font=\tiny, inner sep=0.5pt] {$\perp$};
            \draw[thick] (p11) -- ++(0.7cm, 0.7cm) 
            node[midway, font=\tiny, fill=white, inner sep=1pt] {$\{c_{stop}\}$} 
            node[anchor=north west, font=\tiny, inner sep=0.5pt] {$\perp$};
            \draw[thick] (p20) -- ++(-0.7cm, 0.7cm) 
            node[midway, font=\tiny, fill=white, inner sep=1pt] {$\{c_{stop}\}$} 
            node[anchor=north east, font=\tiny, inner sep=0.5pt] {$\perp$};
            \draw[thick] (p21) -- ++(0.7cm, 0.7cm) 
            node[midway, font=\tiny, fill=white, inner sep=1pt] {$\{c_{stop}\}$} 
            node[anchor=north west, font=\tiny, inner sep=0.5pt] {$\perp$};
        
        \end{tikzpicture}
  \end{minipage}
  \caption{The LSTA $\mathcal{A}_{\mathcal{N}, d}$ (right) obtained from a labeled 1-safe Petri net $\mathcal{N}$ (left).}
  \label{fig:Petri_net_and_LSTA}
\end{figure}

All choice sets are singletons, and each state has at
most one transition for each identifier $c_{\mathrm{init}}$,
$c_{\mathrm{stop}}$, $c_t$ for $t\in\mathcal T$, so the automaton
is choice-deterministic. Distinct net transitions use
distinct choices even if their labels coincide.
There are $\O(m)$ states and $\O(m(|\mathcal T|+1))$
transitions and choice incidences.
Thus, the whole construction has polynomial size in
the input description of the nets. An example of this construction is shown in \Cref{fig:Petri_net_and_LSTA}.

\begin{lemma}%
\label{lem:lsta-pn-traces}
For every %
net $N$, it is the case that
\(
\cL(\cA_{N,d})=E_d(\operatorname{Tr}(N)).
\)
\end{lemma}
\begin{proof}
A run $\rho$ simulating a firing sequence represents a marking by keeping identical copies
of its $j$th bit below the routing address $\alpha_j$.
More precisely, after simulating $i$ transitions to reach
$M_i$, it satisfies
\begin{equation}
\src(\rho(\alpha_j u))=q_{j,M_i(p_j)}
\quad\text{for all }
1\leq j\leq m,\ u\in\{0,1\}^i.
\label{eq:marking-invariant}
\end{equation}
Initialization establishes this invariant for $i=0$.

Consider a transition $t$ enabled at a reachable marking
$M_i$. Every place satisfies the input-token condition
of~\cref{eq:net-simulation}, and 1-safety ensures that
all updated counts belong to $\{0,1\}$.
Thus, every node at this level has its $c_t$ rule.
Using these rules emits $\lambda(t)$ throughout the level
and establishes the invariant for the successor marking.
Any finite firing sequence $t_1\cdots t_k$ therefore gives
an accepting run on $E_d(\lambda(t_1)\cdots\lambda(t_k))$
by taking $c_{\mathrm{stop}}$ after its last step.

Conversely, initialization puts every accepting run in the
states encoding $M_0$.
At each internal level from depth $d$ onward, synchronization
forces all singleton choice sets to be $\{c_t\}$ for one net transition $t$.
Every place has a component, so the existence of these rules
implies that all input places of $t$ contain a token.
Hence $t$ is enabled, and~\cref{eq:net-simulation}
updates every component to the successor marking.
Induction extracts a legal firing sequence $t_1\cdots t_k$.
Its labels and the final stop choice give the tree
$E_d(\lambda(t_1)\cdots\lambda(t_k))$. \qed
\end{proof}

\begin{proof}[of \cref{thm:lsta-inclusion-expspace}]
The two automata use the same map $E_d$.
Its injectivity and~\cref{lem:lsta-pn-traces} give
\(
\operatorname{Tr}(N_1)\subseteq\operatorname{Tr}(N_2)
\iff
\cL(A_{N_1,d})\subseteq\cL(A_{N_2,d}).
\)
The preceding polynomial-time reduction and \Cref{lem:lsta-pn-traces}
establish hardness,
even with one initial state per automaton, singleton
choice sets, one leaf symbol, and labels that are constant at each tree level.

For membership, we use the \EXPSPACE{} upper bound of
Abdulla et al.~\cite[Thm.~15]
{DBLP:journals/pacmpl/AbdullaCCHLLLT25}, which allows
arbitrary binary trees. To restrict acceptance to perfect trees, in both automata we
replace each %
choice $c$ by $2c$ on
internal transitions and by $2c+1$ on leaf transitions.
States, initial states, symbols, and ordered children are unchanged.
Call the resulting automaton $\cA^\sharp$.
An accepting run cannot mix leaf and internal transitions at a level, because their choices have different parity. Thus,
\(
\cL_{\mathrm{original}}(\cA^\sharp)
=\cL_{\mathrm{perfect}}(\cA).
\)
Moreover, this preserves choice determinism and has polynomial
encoding size. \qed
\end{proof}

\section{Inherent growth and sensitivity to qubit order}
\label{sec:representation-growth}

We study unavoidable representation growth under quantum-gate applications and changes of qubit order.
The proofs count distinct subtrees at one level.

\paragraph{Amplitude trees.}
Fix a qubit order $\pi=(v_1,\ldots,v_n)$.
For an amplitude function $f$, write $f_\pi\colon\{0,1\}^n\to\mathbb C$ for its representation in this order, and $T_f^\pi$ for its perfect binary tree: internal depth $i$ is labeled $v_{i+1}$, and leaf $\vec{b}$ is labeled $f_\pi(\vec{b})$.
We keep amplitudes exactly, including global phase.
A \emph{block} $B$ is a finite sequence of gates and acts by $B(T_f^\pi)=T_{Bf}^\pi$, where $Bf$ is the transformed amplitude function.
A \emph{single gate} is a block of length one.
For a layered LSTA $\mathcal{A}$ with order $\pi$, internal transitions at depth $i$ have label $v_{i+1}$.

\paragraph{Residuals and subtrees.}
Fixing a prefix $\vec{a}\in\{0,1\}^i$ leaves the \emph{residual function}
\[
f_{\vec{a}}^\pi(\vec{b})=f_\pi(\vec{a}\vec{b})
\qquad (\vec{b}\in\{0,1\}^{n-i}),
\]
where juxtaposition denotes concatenation.
It gives the leaf labels of the subtree below $\vec{a}$.
For convenience, we sometimes display it as the tuple of its values, listing suffixes $\vec{b}$ in lexicographic order.
At a fixed depth, the subtrees are perfect and their internal labels are fixed by $\pi$, so two residuals are equal exactly when their subtrees are equal.
Let
\[
D_i^\pi(f)=\left|\left\{f_{\vec{a}}^\pi:\vec{a}\in\{0,1\}^i\right\}\right|.
\]
Residuals retain their exact values without normalization.

\begin{lemma}%
\label{lem:residual-lower-bound}
If an LSTA $\cB$ accepts $T_f^\pi$, then
$
|Q_{\cB}|\geq\max_{0 \leq i \leq n} D_i^\pi(f)
$.
\end{lemma}
\begin{proof}
Consider an accepting run on tree $T_f^\pi$.
At depth $i$, distinct residuals give distinct subtrees.
Their roots carry distinct states by \cref{lem:equal-state-subtrees}.
This gives $|Q_{\cB}|\geq D_i^\pi(f)$ at every depth $i$.\qed
\end{proof}
Note that the previous bound applies to every LSTA accepting the tree, even if it accepts other trees or has several accepting runs.

To derive an upper bound, we construct an LSTA $\cB=(Q,I,\Sigma,\Delta)$ accepting only $T_f^\pi$.
Create one state $q_{i,h}$ for each distinct residual $h$ at each depth $i$.
For $i<n$, its two \emph{cofactors} are $h_b(\vec{u})=h(b\vec{u})$, for $b\in\{0,1\}$.
Use the internal and leaf transitions
\[
q_{i,h}\xrightarrow{\{0\}}v_{i+1}(q_{i+1,h_0},q_{i+1,h_1}),
\qquad q_{n,h}\xrightarrow{\{0\}}h(\varepsilon),
\]
with initial state $q_{0,f_\pi}$.
These unique transitions recursively generate exactly $T_f^\pi$.
The state, transition, and choice-incidence counts are equal:
\[
|Q|=|\Delta|=|\inc(\cB)|=\sum_{i=0}^nD_i^\pi(f),\qquad
|\cB|=3\sum_{i=0}^nD_i^\pi(f).
\]
Note that the general lower bound  uses the maximum over depths, while this layered singleton construction uses their sum.
This is the unweighted cofactor DAG of a multi-terminal ordered binary decision diagram (MTBDD), with every variable level retained, and equal residuals merged within each level~\cite{Bryant1986Graphs,DBLP:journals/fmsd/FujitaMY97}.
Reduced MTBDDs may skip irrelevant variable levels.
By \cref{lem:residual-lower-bound}, counting distinct cofactors at a fixed cut also lower-bounds the state count of every LSTA accepting the corresponding tree.

\subsection{Quadratic growth for some quantum gates}

We first give an input quantum state whose total number of residuals is linear in a parameter $m$.
The construction above therefore gives a singleton LSTA of size $\O(m)$.
For our lower bound, we argue that a gate will produce $\Omega(m^2)$ distinct residuals at one depth, forcing that many states in every output representation in the same order (see \Cref{lem:residual-lower-bound}).

\paragraph{A small selector input.}
Fix an integer $k\geq1$ and put $m=2^k$.
A \emph{register} here is a group of $k$ qubits, whose valuation is a $k$-bit address.
We use three registers with valuations $\vec{y},\vec{z},\vec{w}\in\{0,1\}^k$.
For an address $\vec{a}$, let $E_{\vec{a}}(\vec{w})=[\vec{w}=\vec{a}]$.
Its amplitude tuple is the computational basis vector $\ket{\vec{a}}$.
With a selector bit $t$ and $\lambda=1/(\sqrt{2}\,m)$, define
\begin{equation}
f(t,\vec{y},\vec{z},\vec{w})=\lambda\bigl([t=0]E_{\vec{y}}(\vec{w})+[t=1]E_{\vec{z}}(\vec{w})\bigr).
\label{eq:selector-state}
\end{equation}
The selector chooses whether to compare $\vec{w}$ with $\vec{y}$ or with $\vec{z}$.
The unselected address has no influence on that residual.
These are the two root subtrees of one amplitude tree.
There are $2m^2$ nonzero amplitudes, each equal to $\lambda$,
so the sum of their squared moduli is $2m^2|\lambda|^2=1$.
The argument list in \cref{eq:selector-state} groups registers for readability, while the tree uses the interleaved order
\[
\pi_{\mathrm{sel}}=(t,y_1,z_1,\ldots,y_k,z_k,w_1,\ldots,w_k).
\]
Each register has $k$ bits and $m$ addresses; the input has $1+3k$ qubits.

Let $\cA_k$ be the canonical singleton LSTA for $T_f^{\pi_{\mathrm{sel}}}$.
Before the $\vec{w}$ register, a residual remembers the prefix of $\vec{y}$ when $t=0$, or of $\vec{z}$ when $t=1$.
Before $y_i$ there are at most $2^i$ residuals, and before $z_i$ at most $2^i+2^{i-1}=3\cdot2^{i-1}$.
Before $w_1$ there are exactly the $m$ residuals $\lambda E_{\vec{a}}$.
After $j\geq1$ bits of $\vec{w}$ have been read, the residual is zero or a scaled equality function on the remaining $k-j$ bits, giving at most $2^{k-j}+1$ possibilities.
Including the root and leaf levels, these bounds yield
\[
\begin{aligned}
\sum_{\ell=0}^{1+3k}D_\ell^{\pi_{\mathrm{sel}}}(f)
&\leq 1+5(m-1)+m+\sum_{j=1}^k(2^{k-j}+1) =7m+k-5=\O(m).
\end{aligned}
\]
The $m$ distinct residuals before $w_1$ give the matching lower bound.
Every state has one transition with choice set $\{0\}$, so
\begin{equation}
|\cA_k|=3|Q_{\cA_k}|\in\Theta(m).
\label{eq:selector-size}
\end{equation}

\paragraph{Hadamard growth.}

The Hadamard gate replaces each pair of corresponding amplitudes by their normalized sum and difference:
\[
H=\frac1{\sqrt2}\begin{pmatrix}1&1\\1&-1\end{pmatrix}.
\]
With $\lambda=1/(\sqrt2\,m)$ as above, put $\alpha=\lambda/\sqrt2=1/(2m)$.
Applying $H$ to $t$ gives
\[
g(t,\vec{y},\vec{z},\vec{w})=\alpha\bigl(E_{\vec{y}}(\vec{w})+(-1)^tE_{\vec{z}}(\vec{w})\bigr).
\]
Output branch $0$ adds corresponding leaves; branch $1$ subtracts them.
The input and output LSTAs for $k=2$ are shown in \cref{app:quadratic-growth-example}.

\begin{proposition}%
\label{prop:hadamard-growth}
Any LSTA accepting $T_g^{\pi_{\mathrm{sel}}}$ has $\Omega(|\cA_k|^2)$ states.
Moreover, a singleton LSTA for this tree exists with size $\O(|\cA_k|^2)$.
\end{proposition}
\begin{proof}
Consider the cut after $t,y_1,z_1,\ldots,y_k,z_k$ and before $w_1$, at depth $1+2k$.
Prefixes with output $t=0$ give residuals $\alpha R_{\vec{y},\vec{z}}$, where $R_{\vec{y},\vec{z}}=E_{\vec{y}}+E_{\vec{z}}$.
Swapping the addresses leaves this function unchanged.
Two different addresses give two entries equal to $1$, while a repeated address gives one entry equal to $2$.
Thus distinct unordered pairs, allowing repetitions, give different residuals.
On the input's $t=0$ branch, only $\vec{y}$ matters, whereas after Hadamard both $\vec{y}$ and $\vec{z}$ matter.
Counting distinct-address pairs and then repeated addresses gives
\[
\binom m2+\binom m1=\binom{m+1}{2}=\frac{m(m+1)}2.
\]
These are the residuals from the output-$0$ subfamily, so
$D_{1+2k}^{\pi_{\mathrm{sel}}}(g)\geq\binom{m+1}{2}$.
By \cref{lem:residual-lower-bound} and $|\cA_k|\in\Theta(m)$, every accepting LSTA has $\Omega(|\cA_k|^2)$ states.

For the upper bound, propagate pairs of states from the two old root subtrees, together with the output sign.
Each pair carries the two old residuals whose leaves must be added or subtracted and divided by $\sqrt2$.
Since each old state has one transition, at most $\O(|Q_{\cA_k}|^2)$ signed pairs are needed, each with one transition and choice $0$.
Thus, the size belongs to $\O(|\cA_k|^2)$.
This is the singleton instance of the gate construction in~\cite[Thm.~8]{DBLP:journals/pacmpl/AbdullaCCHLLLT25}.\qed
\end{proof}

\paragraph{Backward CX growth.}
The controlled-NOT gate $\mathrm{CX}_{a\to b}$ flips target bit $b$ exactly when control bit $a$ is $1$, leaving all other bits unchanged.
Equivalently, it replaces $b$ by $b\oplus a$, where $\oplus$ denotes exclusive OR.
A controlled gate is \emph{backward} in a qubit order when its
control follows its target.

Add a fresh control qubit $c$, independent of the selector input,
in state $\ket{+}=(\ket{0}+\ket{1})/\sqrt2$.
Place it after the interleaved $\vec{y},\vec{z}$ block:
\[
\pi_{\mathrm{CX}}
=(t,y_1,z_1,\ldots,y_k,z_k,c,w_1,\ldots,w_k).
\]
There are now $2+3k$ qubits, with amplitudes
\[
F(t,\vec{y},\vec{z},c,\vec{w})
=\frac{\lambda}{\sqrt2}
  \bigl([t=0]E_{\vec{y}}(\vec{w})
       +[t=1]E_{\vec{z}}(\vec{w})\bigr).
\]
The amplitudes for $c=0$ and $c=1$ coincide because the two
coefficients of $\ket{+}$ are equal.
The nonzero amplitude is now $\lambda/\sqrt2=\alpha=1/(2m)$.
Relative to the selector residual count, inserting $c$ before $w_1$ adds one level with $m$ residuals.
The uniform change of nonzero scale does not change which residuals are equal.
The canonical construction therefore gives a singleton LSTA
$\widetilde{\cA}_k$ with $|\widetilde{\cA}_k| \in \Theta(m)$.

We now apply $\mathrm{CX}_{c\to t}$ to this input.
Before the gate, fixing $t$ selects which of the two addresses
$\vec{y},\vec{z}$ the residual must remember.
After the gate, this selection depends on $t\oplus c$.
Since $c$ follows $t$ and both address registers in our qubit order,
the output LSTA must distinguish residuals retaining both addresses
until $c$ is read.
The output amplitudes are
\[
G(t,\vec{y},\vec{z},c,\vec{w})=F(t\mathbin\oplus c,\vec{y},\vec{z},c,\vec{w}).
\]

\begin{proposition}%
\label{prop:backward-cx-growth}
Any LSTA accepting $T_G^{\pi_{\mathrm{CX}}}$ needs $\Omega(|\widetilde{\cA}_k|^2)$ states.
Moreover, a singleton LSTA for this tree exists with size $\O(|\widetilde{\cA}_k|^2)$.
\end{proposition}
\begin{proof}
Consider the cut just before $c$, with output $t=0$ and
the addresses $\vec{y},\vec{z}$ fixed.
For $c=0$, this output came from input $t=0$.
For $c=1$, it came from input $t=1$.
The residual is therefore
\[
h_{\vec{y},\vec{z}}(c,\vec{w})=\begin{cases}\alpha E_{\vec{y}}(\vec{w}) &\text{if } c=0,\\\alpha E_{\vec{z}}(\vec{w}) & \text{if } c=1.\end{cases}
\]
Its subtree has root $c$ and the two residuals shown below:
\[
\begin{array}{ccc}
&c&\\
0\,\swarrow&&\searrow\,1\\
\alpha E_{\vec{y}}&&\alpha E_{\vec{z}}
\end{array}
\qquad\longleftrightarrow\qquad
\alpha(E_{\vec{y}}\,\Vert\,E_{\vec{z}}).
\]
Here, $\Vert$ concatenates the value tuples, with the $c=0$ half first and the $c=1$ half second.
For $k=2$, $\vec{y}=01$, and $\vec{z}=10$, this is $\alpha(0,1,0,0\mid0,0,1,0)$.
The first half identifies $\vec{y}$ and the second identifies $\vec{z}$, so there are $m^2$ distinct residuals, including pairs with equal addresses.
The lower bound follows from \cref{lem:residual-lower-bound} and $|\widetilde{\cA}_k|\in\Theta(m)$.

For the upper bound, before $c$, a residual is determined by the output bit $t$ and the prefixes read from both addresses.
After $j$ complete pairs of bits, there are at most $2\cdot4^j$ possibilities and at most twice as many between the two bits of a pair.
Their geometric sum belongs to $\O(m^2)$.
After $c$, only a scaled equality function remains, with $\O(m)$ residuals over the remaining levels as counted above.
The canonical construction gives a singleton representation of size $\O(|\widetilde{\cA}_k|^2)$, matching~\cite[Thms.~8--9]{DBLP:journals/pacmpl/AbdullaCCHLLLT25}.\qed
\end{proof}
Hadamard's output-$(t=0)$ residuals determine unordered pairs with repetition.
The corresponding CX residuals determine ordered pairs, giving $\binom{m+1}{2}$ and $m^2$ residuals, respectively.

For a one-qubit unitary $U$, the controlled gate $\mathrm{C}U_{c\to t}$ applies $U$ to $t$ when $c=1$ and acts as the identity when $c=0$.
\begin{corollary}%
\label{cor:backward-unitary-growth}
For every fixed non-diagonal one-qubit unitary $U$, \cref{prop:backward-cx-growth} also holds for $\mathrm{C}U_{c\to t}F$ in qubit order $\pi_{\mathrm{CX}}$.
\end{corollary}
\begin{proof}
Write $U=(u_{ab})_{a,b\in\{0,1\}}$.
At the same cut with output $t=0$, the residual is
\[
h^U_{\vec{y},\vec{z}}(c,\vec{w})=\begin{cases}
\alpha E_{\vec{y}}(\vec{w}) &\text{if } c=0,\\
\alpha\left(u_{00}E_{\vec{y}}(\vec{w})+u_{01}E_{\vec{z}}(\vec{w})\right)& \text{if } c=1.
\end{cases}
\]
Equality of two residuals gives $\vec{y}=\vec{y}'$ from their $c=0$ halves.
Subtracting their $c=1$ halves then gives $u_{01}(E_{\vec{z}}-E_{\vec{z}'})=0$.
A non-diagonal $2\times2$ unitary has $u_{01}\ne0$, so $\vec{z}=\vec{z}'$.
The same $m^2$ residuals give the lower bound.
The controlled-gate construction in~\cite[Thms.~8--9]{DBLP:journals/pacmpl/AbdullaCCHLLLT25} gives the singleton size upper bound.\qed
\end{proof}

\paragraph{Qubit count.}
The selector uses $1+3k$ qubits and size $\Theta(2^k)$.
Appending $m=2^k$ qubits fixed to zero gives $N=m+1+3k \in \Theta(m)$ qubits and input size $\Theta(N)$: the two input leaf values extend to product-state suffixes of total size $\O(m)$.
Extending every residual by this fixed nonzero suffix is injective, so the Hadamard output still requires $\Omega(N^2)$ states and admits a singleton representation of size $\O(N^2)$.
The same padding gives $N=m+2+3k \in \Theta(m)$ qubits and input
size $\Theta(N)$ for the controlled-gate family.
The residuals used for CX and every fixed non-diagonal
controlled-$U$ remain distinct, giving $\Omega(N^2)$ output states.
Their constantly many leaf values extend through the suffix
with $\O(m)$ additional states, so the singleton upper bound
remains $\O(N^2)$.

\subsection{Qubit-order sensitivity and reordering}
\label{subsec:variable-order}
Variable-order sensitivity is classical for ordered binary decision diagrams (OBDDs)~\cite{Bryant1986Graphs,KrauseSavickyWegener1999}.
By \cref{lem:residual-lower-bound}, lower bounds obtained from distinct cofactors at one cut also apply to LSTAs accepting the tree, even when they accept other trees.

Our selector gives a direct illustration.
Moving $t$ after the interleaved $\vec{y},\vec{z}$ block yields residuals $\lambda(E_{\vec{y}}\Vert E_{\vec{z}})$ before $t$.
There are $m^2$ such residuals, while the input has size $\Theta(m)$.
The same prefix-pair construction as for backward CX gives a matching $\O(m^2)$ singleton upper bound.

For exponential growth, fix an integer $r\geq1$ and use the normalized phase version of the classical inner-product function~\cite{KrauseSavickyWegener1999}:
\[
\psi(\vec{a},\vec{b})=2^{-r}(-1)^{\sum_{j=1}^r a_jb_j},
\qquad \vec{a},\vec{b}\in\{0,1\}^r.
\]
In interleaved order $a_1,b_1,\ldots,a_r,b_r$, remembering the sign and one pending bit gives a singleton LSTA of size $\O(r)$.
In grouped order $a_1,\ldots,a_r,b_1,\ldots,b_r$, the $2^r$ residuals obtained by fixing $\vec{a}$ are distinct.
Two different prefixes are distinguished by a suffix whose only $1$ is at a coordinate where they differ.
Thus, every LSTA accepting this ordered tree needs at least $2^r$ states.

BDD packages have long supported dynamic variable
reordering~\cite{Rudell1993Ordering}.
To reorder an LSTA language, the local exchange of cofactors must also preserve level synchronization.
The following construction provides this adaptation.

Let $\cA=(Q,I,\Sigma,\Delta)$ be a layered LSTA in a fixed $n$-qubit order $\pi$.
For every $0\leq i<n-1$, let $R_i\pi$ exchange the qubits at adjacent internal levels $i,i+1$, and put $R_iT_f^\pi=T_f^{R_i\pi}$.
Write $\Delta_i$ for the transitions at the upper swapped level.
\begin{proposition}[Adjacent reordering]
\label{prop:adjacent-reorder}
An LSTA $R_i(\cA)$ satisfying
\[
\cL(R_i(\cA))=\{R_iT:T\in\cL(\cA)\}
\]
can be constructed in time $\O(|\cA|+|\Delta_i|\,|K(\cA)|)$.
\end{proposition}
\begin{proof}
Put $u=v_{i+1}$ and $v=v_{i+2}$ for the two qubit labels.
For each $\delta:q\xrightarrow C u(p,r)$, create fresh states $s_{\delta,0},s_{\delta,1}$ and replace the transition by
$q\xrightarrow C v(s_{\delta,0},s_{\delta,1})$.
For each pair of transitions $p\xrightarrow D v(p_0,p_1)$ and $r\xrightarrow E v(r_0,r_1)$ with $D\cap E\ne\emptyset$, add
\[
s_{\delta,b}\xrightarrow{D\cap E}u(p_b,r_b)\qquad(b=0,1).
\]
Remove the old states at level $i+1$ and their transitions, which are now unreachable.
All retained transitions respect the new qubit order.

In an original run, the common upper choice still selects $\delta$, and the common lower choice selects a compatible pair of lower transitions.
Exchanging their four ordered cofactors therefore gives a run on the reordered tree with the same choices at the same depths.
Conversely, a common lower choice in a new run determines that pair uniquely by choice determinism.
Both states $s_{\delta,0},s_{\delta,1}$ must therefore use the same pair, reconstructing compatible original transitions.
The other levels are unchanged, proving language equivalence.
A lower choice selects at most one pair for each $\delta$, so the new outgoing choice sets are disjoint.

Index the old transitions by source and choice.
For each $\delta$ and each lower choice, look up the two transitions and group choices selecting the same pair to obtain its intersection set.
This generates at most two choice incidences per $(\delta,\text{choice})$.
Writing $\cA'=R_i(\cA)$, the state, transition, choice-incidence, and choice counts satisfy
\[
\begin{aligned}
|Q_{\cA'}|&\leq |Q|+2|\Delta_i|,&
|\Delta_{\cA'}|&\leq |\Delta|+2|\Delta_i|\,|K(\cA)|,\\
|\inc(\cA')|&\leq |\inc(\cA)|+2|\Delta_i|\,|K(\cA)|,&
|K(\cA')|&\leq |K(\cA)|.
\end{aligned}
\]
Indexing, construction, and removal of obsolete states give the time bound.\qed
\end{proof}
For fixed $\ell_0$, performing at most $\ell_0$ adjacent swaps in the qubit order requires $\O(|\cA|(1+|K(\cA)|)^{\ell_0})$ operations and produces at most that many records in the same model.
A single adjacent swap takes linear time when $|K(\cA)|$ is bounded by a constant.

\section{Discussion and future work}
\label{sec:conclusion}
\label{sec:forward-memory}

The preceding bounds motivate applying circuit blocks with controlled representation growth.

A \emph{monomial} gate maps each computational basis vector to another multiplied by a phase.
Single-qubit examples have diagonal or antidiagonal matrices.
Some monomial gates can be applied by scanning the qubits in order and remembering information about earlier bits.
For example, a CX gate whose control precedes its target only requires remembering the control bit until the target is reached.
For a block of such gates, the scan combines the information retained by its gates and tracks the accumulated phase.
We propose a new measure, \emph{forward-memory width}, to capture this intuition: how much memory this scan uses to distinguish its reachable configurations. The intended gate-application algorithm would then pair each state of a layered LSTA with the scan's possible memory configurations, so bounded width gives only a bounded number of copies of each state.

A verifier could apply blocks of small forward-memory width and use existing gate routines for the remaining gates.
Between blocks, it could try a bounded number of adjacent swaps in the qubit order to place controls before their targets or reduce the information that later blocks must retain, accounting for the reordering cost of \cref{prop:adjacent-reorder}.
For unitary circuits, one may apply leading blocks to the precondition or adjoints (conjugate transposes) of trailing blocks to the postcondition. The adjoint \emph{reverses} the block, moving the postcondition backward and giving the verifier an alternative sequence of intermediate representations.
Preliminary results indicate that this approach is sound (see \cref{app:forward-memory}).
Developing an efficient implementation that combines block application, local reordering, and evaluation from either end of the circuit remains future work, together with assessing whether it broadens the applicability of LSTA-based verification.

\paragraph{Open questions.}
Can equivalent circuits be rewritten to reduce forward-memory width?
Existing quantum-simulation algorithms exploit treewidth and rank-width~\cite{MarkovShi2008,deColnetEtAl2026SumsOfPowers,KuyanovKissinger2026}.
We ask whether finding an order and a partition minimizing the number of gates outside width-$w$ blocks is fixed-parameter tractable in $w$ and a suitable circuit-graph width.
The acyclic $\PiTwoP$ inclusion bound also suggests quantified Boolean formula encodings and, ultimately, circuit synthesis.

Functional inclusion for SWTAs is \PSPACE-hard and in \EXPSPACE~\cite[Thms.~5.6--5.7]{AbdullaEtAl2026Parameterized}.
Our reduction does not directly transfer: functional inclusion compares outputs on the same label-and-color sequence, whereas a trace may be realized using different transition-choice sequences in the two nets.

\bibliographystyle{splncs04}
\bibliography{refs}

\clearpage
\appendix

\section{Polynomial run certificates}
\phantomsection
\label[appendix]{app:acyclic-certificates}

\begin{proof}[of
\cref{lem:acyclic-run-certificate}]
Let $\cA=(Q,I,\Sigma,\Delta)$ be acyclic.
No state repeats along a branch of a run, so every accepted tree has height $h<|Q|$.
Guess a certificate $(q,h,c_0,\ldots,c_h)$ with $q\in I$, $0\leq h<|Q|$, and $c_0,\ldots,c_h\in K(\cA)$.

Starting with $S_0=\{q\}$, for each $0\leq i\leq h$ and $p\in S_i$, find the unique transition $\delta_{p,i}$ from $p$ whose choice set contains $c_i$.
Reject if none exists.
For $i<h$, require all selected transitions to be internal and set
\[
S_{i+1}=\bigcup_{p\in S_i}\trg(\delta_{p,i}).
\]
At $i=h$, require all selected transitions to be leaf transitions.
Construct $D$ with root $(0,q)$ and one node $(i,p)$ per reached pair, labeled by the symbol of $\delta_{p,i}$.
If this transition has ordered child states $(p_0,p_1)$, give the node the ordered children $((i+1,p_0),(i+1,p_1))$.

An accepting run supplies a certificate, and choice determinism ensures that verification recovers its transition at each reached pair $(i,p)$.
Conversely, the transitions recorded during verification of a certificate unfold to an accepting run on $\operatorname{unfold}(D)$: the root state is initial, $c_i$ witnesses synchronization at level $i$, and the arity checks ensure a perfect tree of height $h$.
These checks also cover height $0$ and reject mixed arities at a level.
To conclude, note that there are at most $|Q|$ levels and $|Q|$ states per level, so $D$ has at most $|Q|^2$ nodes.
The certificate has polynomial bit length, and the verification and DAG construction take polynomial time by scanning the explicit transitions and choice sets.
\qed
\end{proof}

\section{Language inclusion for acyclic LSTAs}
\phantomsection
\label[appendix]{app:acyclic-inclusion}
\begin{proof}[of \cref{thm:acyclic-inclusion}]
For noninclusion, guess and verify a run certificate of $\cA$ and construct its tree DAG $D$ using \cref{lem:acyclic-run-certificate}.
Pair each node of $D$ with a state of $\cB$ and retain transitions matching its label and ordered children, with the original choice sets of $\cB$.
Initial states pair the root with $I_{\cB}$.
This polynomial-size product $\cB\!\upharpoonright_D$ is acyclic and satisfies
$\cL(\cB\!\upharpoonright_D)=\cL(\cB)\cap\{\operatorname{unfold}(D)\}$.
Use the acyclic-nonemptiness oracle and accept exactly when it answers empty.
Thus, noninclusion is in $\NP^{\NP}=\SigmaTwoP$.

For hardness, reduce from $\forall\exists$\,3-SAT~\cite{DockerEtAl2020Quantified}.
Let $\forall X\,\exists Y\,\varphi(X,Y)$ be an instance
with $m\geq1$ clauses, on an ordered variable list
$z_1,\ldots,z_N$ with $N\geq1$ and the variables of $X$
first. Put $V=\{1,\ldots,|X|\}$.

Let $\theta$ be the conjunction of $m$ copies of
$z_1\lor\neg z_1$, regarded as a formula on the same
complete ordered variable list, including unused variables.
Apply \cref{lem:assignment-projection} to obtain
$\cA=\mathcal S_{\theta,V}$ and
$\cB=\mathcal S_{\varphi,V}$.
Since both formulas have the same number of clauses
and the same variable list, the two constructions
use the same routing depth $d=\lceil\log_2m\rceil$
and the same height $d+N$.
Thus, their encodings agree at all routing and variable levels,
including the hidden levels for $Y$.

Identify assignments to $X$ with their indexed
assignments in $\{0,1\}^V$.
By \cref{eq:assignment-projection},
\[
\begin{aligned}
\cL(\cA)&=\left\{T_x^V:x\in\{0,1\}^V\right\},\\
\cL(\cB)&=\left\{T_x^V:x\in\{0,1\}^V,
     \ \exists y\in\{0,1\}^Y\ \varphi(x,y)\right\}.
\end{aligned}
\]
Consequently,
\[
\cL(\cA)\subseteq\cL(\cB)
\iff
\forall x\in\{0,1\}^X\;
\exists y\in\{0,1\}^Y:\varphi(x,y).
\]
Both automata have polynomial size, one initial state,
two choices, and the fixed ranked alphabet of
\cref{lem:assignment-projection}.\qed
\end{proof}

\section{Forward-memory blocks and scheduling}
\label[appendix]{app:forward-memory}
In this appendix, we use a RAM model with unit-cost complex arithmetic.

Some gates can be applied by scanning the qubits in order and remembering earlier bits.
For a block of such gates, we bound LSTA growth by the number of memory configurations reachable during this scan.

Fix a qubit order $\pi=(v_1,\ldots,v_n)$.
We represent $n$-qubit vectors by height-$n$ trees with label $v_{i+1}$ at internal level $i$ and amplitudes at the leaves.
An LSTA $\cA=(Q,I,\Sigma,\Delta)$ can be restricted to these trees by pairing states with depths and retaining matching labels, at a cost of $\O((n+1)|\cA|)$.
We use this layered form, with state levels $Q_i$, transitions $\Delta_i$ whose sources lie in $Q_i$, and choice incidences $\inc_i(\cA)=\{(\delta,c)\in\inc(\cA):\delta\in\Delta_i\}$.
Amplitudes have finite encodings on which equality and the required gate arithmetic are computable exactly.
The size bounds below count states, transitions, and choice incidences.

\subsection{Gate templates and width}
A unitary is \emph{monomial} if it maps each computational basis vector to another basis vector multiplied by a phase.
Fix a finite group $\Gamma\leq U(1)=\{z\in\mathbb C:|z|=1\}$ of phases, with effective multiplication and conjugation.
For each gate in the library below, choose a template and a paired template for its adjoint, the conjugate transpose of its matrix.
A template has a finite memory set $\mathcal R_i$ after each qubit $v_i$ and initial memory $m_0\in\mathcal R_0$.
At qubit $v_i$ (tree depth $i-1$), it reads an input bit $x$ with memory $m\in\mathcal R_{i-1}$ and uses a rule
\[
(m,x)\mapsto
\bigl(\sigma_{i,m}(x),\delta_i(m,x),\eta_i(m,x)\bigr),
\quad \eta_i(m,x)\in\Gamma.
\]
The three outputs are the output bit, the next memory in $\mathcal R_i$, and a phase factor.
Here $\sigma_{i,m}$ permutes $\{0,1\}$, and the product of the emitted phases multiplies the input amplitude.
It describes a \emph{prefix-causal monomial gate}: the output prefix depends only on the input prefix.
Its inverse reads $y$, recovers $x=\sigma_{i,m}^{-1}(y)$, updates to $\delta_i(m,x)$, and emits $\overline{\eta_i(m,x)}$.
The memory update itself may forget earlier bits.

Our library uses the following templates:
\begin{itemize}
\item Single-qubit monomial gates with nonzero entries in $\Gamma$ need no pending register.
\item Controlled versions whose controls precede the target retain the control predicate, using one partial-AND bit for a conjunction.
\item Constant-arity diagonal gates with phases in $\Gamma$ retain earlier participating bits until the final wire.
\end{itemize}
Each register is discarded after its last use.
Constant-arity prefix-causal monomial gates may be added with fixed templates storing their earlier participating bits.
For a positive integer $r$, write $\mu_r=\{e^{2\pi\mathrm{i}j/r}:j=0,\ldots,r-1\}$ for the $r$th roots of unity.
For $\Gamma=\mu_8$, the library includes the Pauli gates $X,Y,Z$ and phase gates $S,T$, where
\[
\begin{gathered}
X=\begin{pmatrix}0&1\\1&0\end{pmatrix},\qquad
Y=\begin{pmatrix}0&-\mathrm{i}\\\mathrm{i}&0\end{pmatrix},\qquad
Z=\operatorname{diag}(1,-1),\\
S=\operatorname{diag}(1,\mathrm{i}),\qquad
T=\operatorname{diag}(1,e^{\mathrm{i}\pi/4}).
\end{gathered}
\]
It also includes their adjoints and controlled versions with controls before the target.
The diagonal gates $CZ,CCZ,CS,CT$ are supported in any qubit order.
Each $C$ adds a control that must be $1$ for the gate to act.
For $\theta\in\mathbb R$, the library also includes $R_Z(\theta)=\operatorname{diag}(e^{-\mathrm{i}\theta/2},e^{\mathrm{i}\theta/2})$ whenever both diagonal entries belong to $\Gamma$.

For a supported block $B=G_\ell\cdots G_1$, feed each scanned bit through the templates in chronological order, retain their registers, and accumulate emitted phases in one register $\gamma\in\Gamma$.
Here $G_1$ acts first, $|B|=\ell$ is the number of gates, and the phase register starts at $1$.
For $0\leq i\leq n$, let $\mathcal M_i(B,\pi)$ be the configurations reachable on input prefixes in $\{0,1\}^i$.
Each tuple records the gate memories $m_1,\ldots,m_\ell$ and the phase $\gamma$.
This compilation uses the fixed templates and removes unreachable configurations.
For the compiled block, we reuse $\sigma_{i,m}$, $\delta_i$, and $\eta_i$ for its bit permutation, tuple update, and phase factor, with $m=(m_1,\ldots,m_\ell)$.
Its initial memory is the tuple $m_0=(m_{1,0},\ldots,m_{\ell,0})$ of the individual templates' initial memories.
Write $M_i(B,\pi)=|\mathcal M_i(B,\pi)|$, or $M_i$ when the block and order are clear.

\begin{definition}[Syntactic forward-memory width]
\label{def:syntactic-fmw}
The forward-memory width of $B$ in order $\pi$ is
\[
\operatorname{fmw}_\pi(B)=\left\lceil\log_2\max_{0\leq i\leq n} M_i(B,\pi)\right\rceil.
\]
\end{definition}

For each gate occurrence $g$ in $B$, let $\mathcal R_i^g$ be its register's value set, taken as a singleton when inactive.
Then
\[
M_i\leq\widehat M_i:=|\Gamma|\prod_{g\in B}|\mathcal R_i^g|.
\]
Counting reachable configurations can give a smaller bound: fan-out from one control has only two reachable control tuples, and a block of single-qubit $T$ gates accumulates at most eight phases.
Reachable compilation takes $\O(|B|\sum_i M_i)$ operations.
The phase register also contributes to width.
For $\Gamma=\mu_{2^{n+1}}$, applying
\[
\operatorname{diag}(1,e^{2\pi\mathrm{i}/2^{j+1}})\quad\text{to qubit }v_j,
\qquad 1\leq j\leq n,
\]
produces $2^n$ distinct leaf values from $\ket{+}^{\otimes n}$ with no pending controls.

\subsection{Applying blocks and adjoints}

\begin{proposition}
\label{prop:fmw-apply}
For a layered LSTA $\cA$ and a precompiled block $B$ of width $w$ in the same qubit order, one can construct $B(\cA)$ using $\O(2^w|\cA|)$ operations, with
\[
\cL(B(\cA))=\{B(T):T\in\cL(\cA)\}.
\]
\end{proposition}
\begin{proof}
Use states $(q,m,\gamma)$.
For $q\xrightarrow C v_{i+1}(q_0,q_1)$, copy $C$ and give output child $y$ the state
$(q_x,\delta_{i+1}(m,x),\gamma\eta_{i+1}(m,x))$, where $x=\sigma_{i+1,m}^{-1}(y)$.
Replace a leaf rule $q\xrightarrow C a$ by $(q,m,\gamma)\xrightarrow C\gamma a$.
Initial states carry $m_0$ and phase $1$.
The branch permutations give a bijection of runs preserving every common level choice and applying exactly the forward phase to each recovered input amplitude.
Choice determinism follows from copying outgoing choice sets.
At level $i$, each record is copied at most $M_i$ times.
Writing $\cA'=B(\cA)$, we obtain
\[
\begin{aligned}
|Q_{\cA'}|&\leq\sum_i M_i|Q_i|,&
|\Delta_{\cA'}|&\leq\sum_i M_i|\Delta_i|,\\
|\inc(\cA')|&\leq\sum_i M_i|\inc_i(\cA)|,&
|K(\cA')|&\leq|K(\cA)|.
\end{aligned}
\]
These counts give the work bound.
Leaf labels belong to $\Gamma\Sigma_0=\{\gamma a:\gamma\in\Gamma,\ a\in\Sigma_0\}$.\qed
\end{proof}

Write $B^\dagger=G_1^\dagger\cdots G_\ell^\dagger$ for the adjoint block.
\begin{proposition}
\label{prop:fmw-adjoint}
Paired inverse compilation satisfies $M_i(B,\pi)=M_i(B^\dagger,\pi)$ at every cut.
\end{proposition}
\begin{proof}
Each input prefix determines a unique output prefix, and conversely.
Inverse execution has configuration $(m_\ell,\ldots,m_1,\overline\gamma)$, so reversal of the register tuple and phase conjugation give a bijection at every cut.
\qed
\end{proof}
Thus adjoints preserve width and the product estimate $\widehat M_i$.
For composition, configurations are images of pairs of block configurations, with their phases multiplied.
Consequently,
\[
\operatorname{fmw}_\pi(B_2B_1)\leq
\operatorname{fmw}_\pi(B_1)+\operatorname{fmw}_\pi(B_2).
\]

\subsection{A scheduling heuristic}
We call gates outside the template library \emph{separators}, as they separate forward-memory blocks.
They include Hadamards and unsupported backward permutations and are applied by exact gate routines~\cite[Algs.~1--3]{DBLP:journals/pacmpl/AbdullaCCHLLLT25}.
The planner also treats a gate as a separator when its compilation exceeds the allotted budget.
The SWAP gate exchanges two qubit values, and applying it in a fixed qubit order incurs its transformation cost.

For a unitary circuit, the proposed planner maintains two automata in a common qubit order and the remaining chronological gates $G_l,\ldots,G_r$.
The current verification obligation is
\begin{equation}
(G_r\cdots G_l)(\cL(\cA_L))\subseteq\cL(\cA_R).
\label{eq:two-ended-obligation}
\end{equation}
Fix nonnegative integer bounds $b$ on width and $\ell_0$ on swaps, with $\ell_0$ constant, and a positive integer shortlist size $\kappa$.
For initial alignment, the planner compares a supplied finite set of qubit orders containing both endpoint orders and orders derived from the circuit.
It ranks their greedy partitions by separator count, then block count, and then block widths, and retains several for the trials below.
Each trial includes the cost of aligning both endpoints.
At each subsequent boundary, it inspects the $n^{\O(\ell_0)}$ qubit orders within $\ell_0$ adjacent swaps, including the current qubit order.
From each circuit end, it greedily extends a consecutive supported block while reachable compilation stays within the width cap $b$.
The cap is tested on reachable configurations, using the product estimate only as an upper bound.
The ranking prefers supported blocks, then more consumed gates, lower width, and fewer swaps.
It retains $\kappa$ candidates per end together with exact single-gate trials in the current qubit order.

For each shortlisted candidate, the executor tentatively reorders both endpoints, applies language-preserving reduction~\cite[Sect.~6.3]{DBLP:journals/pacmpl/AbdullaCCHLLLT25}, and counts their states, transitions, choices, and choice incidences at each level.
It ranks feasible trials by reordering work plus $\sum_iM_i(|Q_i|+|\Delta_i|+|\inc_i(\cA)|)$ on the chosen endpoint $\cA$, divided by the number of gates consumed.
A trial is rolled back when the peak memory used by the two endpoints together exceeds the limit.
Separators use their exact routine's bound, such as the quadratic product bounds for one-qubit and controlled gates~\cite[Thms.~8--9]{DBLP:journals/pacmpl/AbdullaCCHLLLT25}.
Each iteration commits the common qubit order, applies a nonempty left block $B$ to $\cA_L$ or the adjoint of a nonempty right block $D$ to $\cA_R$, and removes those gates.
When none remain, it tests LSTA inclusion.
Throughout, changes in qubit order preserve physical qubit identities, and adjoints are applied in the current qubit order.

\begin{proposition}
\label{prop:two-ended-soundness}
Every completed schedule returns the answer to the original verification obligation.
\end{proposition}
\begin{proof}
Left application preserves~\cref{eq:two-ended-obligation} by associativity.
Right application uses $D(X)\subseteq Y\iff X\subseteq D^\dagger(Y)$.
Reordering is a bijection on both languages and conjugates the same physical circuit by the coordinate permutation.
Exact transformations and language-preserving reduction therefore preserve the truth of the obligation.
Each iteration consumes gates, so there are at most as many iterations as original gates.\qed
\end{proof}

The ranking is a heuristic for choosing transformations.
Total verification work includes compilation, unsuccessful trials, reduction, and final inclusion, as well as the selected transformations.
If all trials exceed a resource limit, verification remains unresolved.

Intermediate sizes affect this cost: applying two Hadamards separately to the same qubit can produce the quadratic intermediate of \cref{prop:hadamard-growth} before returning to the input.
Taking the adjoint reverses the block widths of a fixed partition.
For input size $s$ and a block of width $w$, put $M=2^w$.
Applying the bounds $s\mapsto Ms$ and $s\mapsto s^2$ in opposite orders gives $M^2s^2$ and $Ms^2$, respectively.
Greedy partitioning in the two directions can also select different blocks.

The local search follows window reordering for binary decision diagrams~\cite{Rudell1993Ordering}.
Retaining several orders and comparing their estimated costs follows cost-based query planning~\cite{SelingerEtAl1979Access}.
Factorized query results~\cite{OlteanuZavodny2012Factorised} also exploit shared structure, while worst-case optimal joins~\cite{NgoEtAl2012Joins} avoid large intermediate products.

Forward-memory width measures the given circuit syntax: even canceling CX gates retain separate registers during compilation.
Exact circuit rewriting, such as cancellation, commutation, or alternative gate decompositions, could lower this width or remove separators.
Minimizing width over equivalent circuits remains a question for future work.

\newpage
\section{Language inclusion for acyclic LSTAs: example}
\label[appendix]{app:acyclic-inclusion-example}
\begin{figure}[H]
\centering
    \begin{minipage}[b]{0.3\textwidth}
    \centering
    \begin{tikzpicture}[
        yscale=-1,
        state/.style={rounded rectangle, draw=black, font=\tiny},
        ]
        \node[state] (q) at (0, 0) {$q$};
        \node[state] (q_1) at (0, 2) {$q_{1}$};
        \node[state] (q_2) at (0, 4) {$q_{2}$};
        \node[state] (q_3) at (0, 6) {$q_{3}$};
        \node[state] (q_4) at (0, 8) {$q_{4}$};

        \draw[-{Stealth[scale=0.9]}] ($(q) + (-0.5cm, 0)$) -- (q);

        \transitionSector{q}{q_1}{q_1}{\{0\}}{$\bullet$}
        \transitionSector[shift={-0.8cm, 0.8cm}]{q_1}{q_2}{q_2}{\{0\}}{$a_0$}
        \transitionSector[shift={0.8cm, 0.8cm}]{q_1}{q_2}{q_2}{\{1\}}{$a_1$}
        \transitionSector[shift={-0.8cm, 0.8cm}]{q_2}{q_3}{q_3}{\{0\}}{$a_0$}
        \transitionSector[shift={0.8cm, 0.8cm}]{q_2}{q_3}{q_3}{\{1\}}{$a_1$}
        \transitionSector{q_3}{q_4}{q_4}{\{0, 1\}}{$\bullet$}
        
        \draw[thick] (q_4) -- ++(0,1cm) node[midway, font=\tiny, fill=white, inner sep=1pt] {\{0\}} node[below, font=\tiny] {\#};
    \end{tikzpicture}
    \end{minipage}
    \hfill
    \begin{minipage}[b]{0.65\textwidth}
    \centering
    \begin{tikzpicture}[
        yscale=-1,
        state/.style={rounded rectangle, draw=black, font=\tiny},
        ]
        \node[state] (q) at (2, 0) {$q$};
        \node[state] (000) at (0, 2) {$1, 0, 0$};
        \node[state] (100) at (4, 2) {$2, 0, 0$};
        \node[state] (010) at (0, 4) {$1, 1, 0$};
        \node[state] (011) at (2, 4) {$1, 1, 1$};
        \node[state] (110) at (4, 4) {$2, 1, 0$};
        \node[state] (111) at (6, 4) {$2, 1, 1$};
        \node[state] (020) at (0, 6) {$1, 2, 0$};
        \node[state] (021) at (2, 6) {$1, 2, 1$};
        \node[state] (120) at (4, 6) {$2, 2, 0$};
        \node[state] (121) at (6, 6) {$2, 2, 1$};
        \node[state] (030) at (0, 8) {$1, 3, 0$};
        \node[state] (031) at (2, 8) {$1, 3, 1$};
        \node[state] (130) at (4, 8) {$2, 3, 0$};
        \node[state] (131) at (6, 8) {$2, 3, 1$};

        \draw[-{Stealth[scale=0.9]}] ($(q) + (-0.5cm, 0)$) -- (q);

        \transitionSector[sector dist=1cm, symbol dist=1cm, apex factor=5]{q}{000}{100}{\{0\}}{$\bullet$}
        \transitionSector[]{100}{110}{110}{\{1\}}{$a_1$}
        \transitionSector[]{000}{011}{011}{\{1\}}{$a_1$}
        \transitionSector[]{000}{010}{010}{\{0\}}{$a_0$}
        \transitionSector[]{100}{111}{111}{\{0\}}{$a_0$}
        \transitionSector[]{110}{121}{121}{\{1\}}{$a_1$}
        \transitionSector[]{110}{120}{120}{\{0\}}{$a_0$}
        \transitionSector[shift={-0.45cm, 0.8cm}]{111}{121}{121}{\{0\}}{\raisebox{7pt}{$a_0$}}
        \transitionSector[shift={0.45cm, 0.8cm}]{111}{121}{121}{\{1\}}{$a_1$}
        \transitionSector[shift={-0.45cm, 0.8cm}]{011}{021}{021}{\{0\}}{\raisebox{7pt}{$a_0$}}
        \transitionSector[shift={0.45cm, 0.8cm}]{011}{021}{021}{\{1\}}{$a_1$}
        \transitionSector[]{010}{020}{020}{\{0\}}{$a_0$}
        \transitionSector[]{010}{021}{021}{\{1\}}{$a_1$}
        \transitionSector[]{021}{031}{031}{\{0, 1\}}{$\bullet$}
        \transitionSector[]{020}{030}{030}{\{0\}}{$\bullet$}
        \transitionSector[]{121}{131}{131}{\{0, 1\}}{$\bullet$}
        \transitionSector[]{020}{031}{031}{\{1\}}{$\bullet$}
        \transitionSector[]{120}{130}{130}{\{1\}}{$\bullet$}
        \transitionSector[]{120}{131}{131}{\{0\}}{$\bullet$}
        
        \draw[thick] (131) -- ++(0,1cm) node[midway, font=\tiny, fill=white, inner sep=1pt] {\{0\}} node[below, font=\tiny] {\#};
        \draw[thick] (031) -- ++(0,1cm) node[midway, font=\tiny, fill=white, inner sep=1pt] {\{0\}} node[below, font=\tiny] {\#};
    \end{tikzpicture}
    \end{minipage}
\caption{LSTAs $\cA$ (left) and $\cB$ (right) such that $\cL(\cA) \subseteq \cL(\cB)$ if and only if it is the case that $\forall x_1, x_2 \ \exists y_1 \; (x_1\lor x_2 \lor y_1) \land (\neg x_1\lor x_2 \lor \neg y_1)$.}
\end{figure}
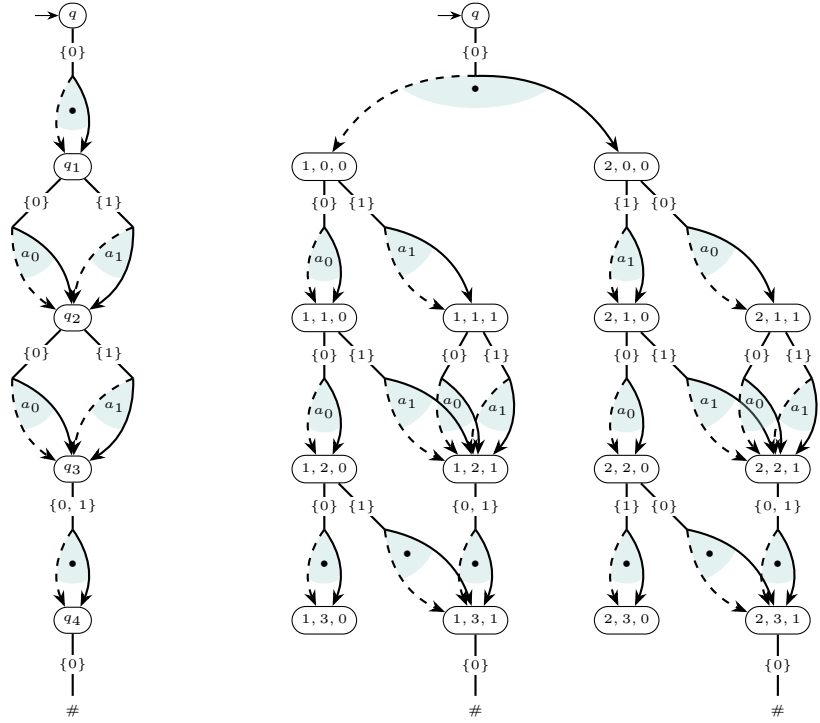

\newpage
\section{Quadratic growth example}
\label[appendix]{app:quadratic-growth-example}
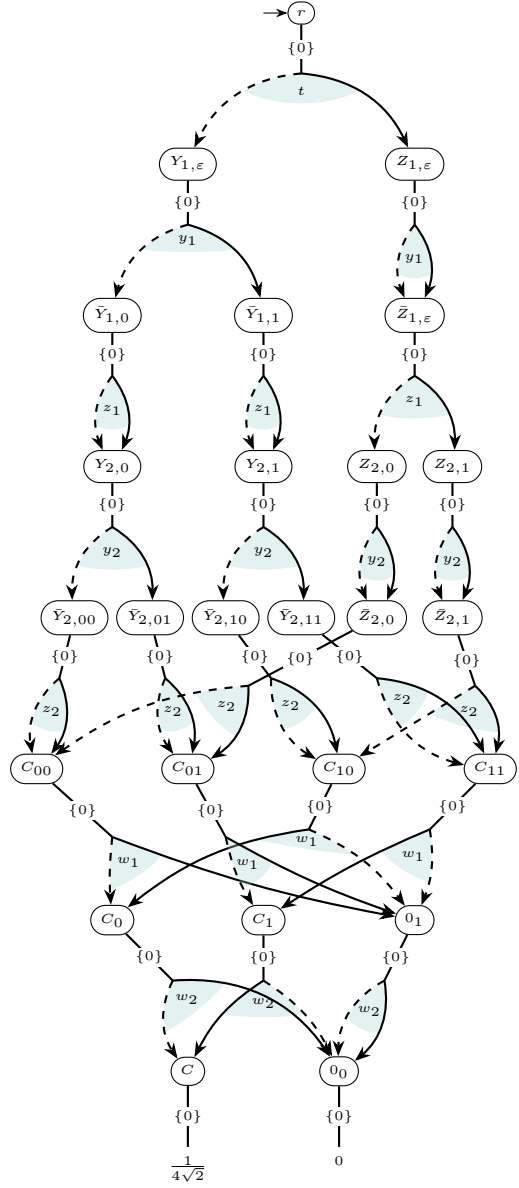
\begin{figure}[H]
\centering
    \begin{tikzpicture}[
        yscale=-1,
        state/.style={rounded rectangle, draw=black, font=\tiny},
        ]
        \node[state] (r) at (0.5, 0) {$r$};
        \node[state] (Y_1_) at (-1, 2) {$Y_{1, \varepsilon}$};
        \node[state] (Z_1_) at (2, 2) {$Z_{1, \varepsilon}$};
        \node[state] (Yb_1_0) at (-2, 4) {$\bar{Y}_{1, 0}$};
        \node[state] (Yb_1_1) at (0, 4) {$\bar{Y}_{1, 1}$};
        \node[state] (Zb_1_) at (2, 4) {$\bar{Z}_{1, \varepsilon}$};
        \node[state] (Y_2_0) at (-2, 6) {$Y_{2, 0}$};
        \node[state] (Y_2_1) at (0, 6) {$Y_{2, 1}$};
        \node[state] (Z_2_0) at (1.5, 6) {$Z_{2, 0}$};
        \node[state] (Z_2_1) at (2.5, 6) {$Z_{2, 1}$};
        \node[state] (Yb_2_00) at (-2.5, 8) {$\bar{Y}_{2, 00}$};
        \node[state] (Yb_2_01) at (-1.5, 8) {$\bar{Y}_{2, 01}$};
        \node[state] (Yb_2_10) at (-0.5, 8) {$\bar{Y}_{2, 10}$};
        \node[state] (Yb_2_11) at (0.5, 8) {$\bar{Y}_{2, 11}$};
        \node[state] (Zb_2_0) at (1.5, 8) {$\bar{Z}_{2, 0}$};
        \node[state] (Zb_2_1) at (2.5, 8) {$\bar{Z}_{2, 1}$};
        \node[state] (C_00) at (-3, 10) {$C_{00}$};
        \node[state] (C_01) at (-1, 10) {$C_{01}$};
        \node[state] (C_10) at (1, 10) {$C_{10}$};
        \node[state] (C_11) at (3, 10) {$C_{11}$};
        \node[state] (C_0) at (-2, 12) {$C_{0}$};
        \node[state] (C_1) at (0, 12) {$C_{1}$};
        \node[state] (O_1) at (2, 12) {$0_{1}$};
        \node[state] (C_) at (-1, 14) {$C_{}$};
        \node[state] (O_0) at (1, 14) {$0_{0}$};

        \draw[-{Stealth[scale=0.9]}] ($(r) + (-0.5cm, 0)$) -- (r);
        
        \transitionSector[sector dist=0.8cm, symbol dist=0.8cm, apex factor=3]{r}{Y_1_}{Z_1_}{\{0\}}{$t$}
        \transitionSector{Y_1_}{Yb_1_0}{Yb_1_1}{\{0\}}{$y_1$}
        \transitionSector{Z_1_}{Zb_1_}{Zb_1_}{\{0\}}{$y_1$}
        \transitionSector{Yb_1_0}{Y_2_0}{Y_2_0}{\{0\}}{$z_1$}
        \transitionSector{Yb_1_1}{Y_2_1}{Y_2_1}{\{0\}}{$z_1$}
        \transitionSector{Zb_1_}{Z_2_0}{Z_2_1}{\{0\}}{$z_1$}
        \transitionSector{Z_2_0}{Zb_2_0}{Zb_2_0}{\{0\}}{$y_2$}
        \transitionSector{Z_2_1}{Zb_2_1}{Zb_2_1}{\{0\}}{$y_2$}
        \transitionSector{Y_2_1}{Yb_2_10}{Yb_2_11}{\{0\}}{$y_2$}
        \transitionSector{Y_2_0}{Yb_2_00}{Yb_2_01}{\{0\}}{$y_2$}
        \transitionSector{Yb_2_00}{C_00}{C_00}{\{0\}}{$z_2$}
        \transitionSector[shift={0.3cm, 0.9cm}, angle left=0]{Zb_2_1}{C_10}{C_11}{\{0\}}{$z_2$}
        \transitionSector{Yb_2_11}{C_11}{C_11}{\{0\}}{$z_2$}
        \transitionSector{Yb_2_10}{C_10}{C_10}{\{0\}}{$z_2$}
        \transitionSector{Yb_2_01}{C_01}{C_01}{\{0\}}{$z_2$}
        \transitionSector[shift={-1.7cm, 0.9cm}, angle left=15]{Zb_2_0}{C_00}{C_01}{\{0\}}{$z_2$}
        \transitionSector[angle left=-15, angle right=-15]{C_10}{O_1}{C_0}{\{0\}}{$w_1$}
        \transitionSector[shift={1cm, 0.9cm}, angle left=5, angle right=-5]{C_00}{C_0}{O_1}{\{0\}}{$w_1$}
        \transitionSector[angle left=-15, angle right=-5]{C_11}{O_1}{C_1}{\{0\}}{$w_1$}
        \transitionSector[shift={0.5cm, 0.9cm}, angle left=10, angle right=-5]{C_01}{C_1}{O_1}{\{0\}}{$w_1$}
        \transitionSector{O_1}{O_0}{O_0}{\{0\}}{$w_2$}
        \transitionSector[angle left=-15, angle right=-15]{C_1}{O_0}{C_}{\{0\}}{$w_2$}
        \transitionSector{C_0}{C_}{O_0}{\{0\}}{$w_2$}
        
        \draw[thick] (O_0) -- ++(0,1cm) node[midway, font=\tiny, fill=white, inner sep=1pt] {\{0\}} node[below, font=\tiny] {$0$};
        \draw[thick] (C_) -- ++(0,1cm) node[midway, font=\tiny, fill=white, inner sep=1pt] {\{0\}} node[below, font=\tiny] {$\frac{1}{4\sqrt{2}}$};
    \end{tikzpicture}
    \caption{The selector LSTA $\cA_2$ for $k=2$, representing $T_f^{\pi_{\mathrm{sel}}}$.
    States $Y_{i,u}$ and $Z_{i,u}$ remember prefixes $u$ of $\vec y$ and $\vec z$, respectively, before $y_i$; barred states do so before $z_i$.
    State $C_u$ represents the residual $\lambda E_u$ on the remaining $\vec w$ bits, with $C_{\varepsilon}$ written $C$, and $0_j$ represents the zero residual with $j$ bits remaining.}
    \label{fig:selector-input}
\end{figure}

\begin{figure}[H]
\centering
    \begin{tikzpicture}[
        yscale=-1,
        x={\dimexpr(\textwidth - 1.5cm)/44\relax},
        state/.style={rounded rectangle, draw=black, font=\tiny},
        ]
        \node[state] (r) at (22, 0) {$r$};

        \node[state] (Y_1_Z_1_L) at (21, 2) {};
        \node[state] (Y_1_Z_1_R) at (23, 2) {};

        \node[state] (Yb_1_0_Zb_1_L) at (19, 4) {};
        \node[state] (Yb_1_1_Zb_1_L) at (21, 4) {};
        \node[state] (Yb_1_0_Zb_1_R) at (23, 4) {};
        \node[state] (Yb_1_1_Zb_1_R) at (25, 4) {};

        \node[state] (Y_2_0_Z_2_0_L) at (15, 6) {};
        \node[state] (Y_2_0_Z_2_1_L) at (17, 6) {};
        \node[state] (Y_2_1_Z_2_0_L) at (19, 6) {};
        \node[state] (Y_2_1_Z_2_1_L) at (21, 6) {};
        \node[state] (Y_2_0_Z_2_0_R) at (23, 6) {};
        \node[state] (Y_2_0_Z_2_1_R) at (25, 6) {};
        \node[state] (Y_2_1_Z_2_0_R) at (27, 6) {};
        \node[state] (Y_2_1_Z_2_1_R) at (29, 6) {};

        \node[state] (Yb_2_00_Zb_2_0_L) at (7, 8) {};
        \node[state] (Yb_2_01_Zb_2_0_L) at (9, 8) {};
        \node[state] (Yb_2_00_Zb_2_1_L) at (11, 8) {};
        \node[state] (Yb_2_01_Zb_2_1_L) at (13, 8) {};
        \node[state] (Yb_2_10_Zb_2_0_L) at (15, 8) {};
        \node[state] (Yb_2_11_Zb_2_0_L) at (17, 8) {};
        \node[state] (Yb_2_10_Zb_2_1_L) at (19, 8) {};
        \node[state] (Yb_2_11_Zb_2_1_L) at (21, 8) {};
        \node[state] (Yb_2_00_Zb_2_0_R) at (23, 8) {};
        \node[state] (Yb_2_01_Zb_2_0_R) at (25, 8) {};
        \node[state] (Yb_2_00_Zb_2_1_R) at (27, 8) {};
        \node[state] (Yb_2_01_Zb_2_1_R) at (29, 8) {};
        \node[state] (Yb_2_10_Zb_2_0_R) at (31, 8) {};
        \node[state] (Yb_2_11_Zb_2_0_R) at (33, 8) {};
        \node[state] (Yb_2_10_Zb_2_1_R) at (35, 8) {};
        \node[state] (Yb_2_11_Zb_2_1_R) at (37, 8) {};

        \node[state] (C_00_C_00_L) at (0, 10) {};
        \node[state] (C_01_C_00_L) at (2, 10) {};
        \node[state] (C_01_C_01_L) at (4, 10) {};
        \node[state] (C_10_C_00_L) at (6, 10) {};
        \node[state] (C_00_C_11_L) at (8, 10) {};
        \node[state] (C_01_C_10_L) at (10, 10) {};
        \node[state] (C_01_C_11_L) at (12, 10) {};
        \node[state] (C_10_C_10_L) at (14, 10) {};
        \node[state] (C_10_C_11_L) at (16, 10) {};
        \node[state] (C_11_C_11_L) at (18, 10) {};
        \node[state] (C_01_C_01_R) at (20, 10) {};
        \node[state] (C_00_C_01_R) at (22, 10) {};
        \node[state] (C_01_C_00_R) at (24, 10) {};
        \node[state] (C_00_C_10_R) at (26, 10) {};
        \node[state] (C_00_C_11_R) at (28, 10) {};
        \node[state] (C_01_C_10_R) at (30, 10) {};
        \node[state] (C_01_C_11_R) at (32, 10) {};
        \node[state] (C_10_C_00_R) at (34, 10) {};
        \node[state] (C_10_C_01_R) at (36, 10) {};
        \node[state] (C_11_C_00_R) at (38, 10) {};
        \node[state] (C_11_C_01_R) at (40, 10) {};
        \node[state] (C_10_C_11_R) at (42, 10) {};
        \node[state] (C_11_C_10_R) at (44, 10) {};

        \node[state] (C_0_C_0_L) at (2, 12) {};
        \node[state] (O_1_O_1_L) at (6.5, 12) {};
        \node[state] (C_1_C_0_L) at (11, 12) {};
        \node[state] (C_1_C_1_L) at (15.5, 12) {};
        \node[state] (O_1_C_0_L) at (20, 12) {};
        \node[state] (C_1_O_1_L) at (24, 12) {};
        \node[state] (C_0_C_1_R) at (28.5, 12) {};
        \node[state] (C_1_C_0_R) at (33, 12) {};
        \node[state] (O_1_C_0_R) at (37.5, 12) {};
        \node[state] (O_1_C_1_R) at (42, 12) {};

        \node[state] (C_C_L) at (6.5, 14) {};
        \node[state] (O_0_O_0_L) at (17, 14) {};
        \node[state] (C_O_0_L) at (27, 14) {};
        \node[state] (O_0_C_R) at (37.5, 14) {};

        \draw[-{Stealth[scale=0.9]}] ($(r) + (-0.5cm, 0)$) -- (r);

        \transitionSector[angle left=5, angle right=5]{r}{Y_1_Z_1_L}{Y_1_Z_1_R}{}{}
        \transitionSector[angle left=5, angle right=5]{Y_1_Z_1_L}{Yb_1_0_Zb_1_L}{Yb_1_1_Zb_1_L}{}{}
        \transitionSector[angle left=5, angle right=5]{Y_1_Z_1_R}{Yb_1_0_Zb_1_R}{Yb_1_1_Zb_1_R}{}{}
        \transitionSector[angle left=5, angle right=5]{Yb_1_1_Zb_1_R}{Y_2_1_Z_2_0_R}{Y_2_1_Z_2_1_R}{}{}
        \transitionSector[angle left=5, angle right=5]{Yb_1_1_Zb_1_L}{Y_2_1_Z_2_0_L}{Y_2_1_Z_2_1_L}{}{}
        \transitionSector[angle left=5, angle right=5]{Yb_1_0_Zb_1_R}{Y_2_0_Z_2_0_R}{Y_2_0_Z_2_1_R}{}{}
        \transitionSector[angle left=5, angle right=5]{Yb_1_0_Zb_1_L}{Y_2_0_Z_2_0_L}{Y_2_0_Z_2_1_L}{}{}
        \transitionSector[angle left=5, angle right=5]{Y_2_0_Z_2_0_R}{Yb_2_00_Zb_2_0_R}{Yb_2_01_Zb_2_0_R}{}{}
        \transitionSector[angle left=5, angle right=5]{Y_2_1_Z_2_1_R}{Yb_2_10_Zb_2_1_R}{Yb_2_11_Zb_2_1_R}{}{}
        \transitionSector[angle left=5, angle right=5]{Y_2_1_Z_2_0_L}{Yb_2_10_Zb_2_0_L}{Yb_2_11_Zb_2_0_L}{}{}
        \transitionSector[angle left=5, angle right=5]{Y_2_0_Z_2_1_R}{Yb_2_00_Zb_2_1_R}{Yb_2_01_Zb_2_1_R}{}{}
        \transitionSector[angle left=5, angle right=5]{Y_2_1_Z_2_0_R}{Yb_2_10_Zb_2_0_R}{Yb_2_11_Zb_2_0_R}{}{}
        \transitionSector[angle left=5, angle right=5]{Y_2_0_Z_2_0_L}{Yb_2_00_Zb_2_0_L}{Yb_2_01_Zb_2_0_L}{}{}
        \transitionSector[angle left=5, angle right=5]{Y_2_1_Z_2_1_L}{Yb_2_10_Zb_2_1_L}{Yb_2_11_Zb_2_1_L}{}{}
        \transitionSector[angle left=5, angle right=5]{Y_2_0_Z_2_1_L}{Yb_2_00_Zb_2_1_L}{Yb_2_01_Zb_2_1_L}{}{}
        \transitionSector[angle left=5, angle right=5]{Yb_2_00_Zb_2_1_L}{C_10_C_00_L}{C_00_C_11_L}{}{}
        \transitionSector[angle left=5, angle right=5]{Yb_2_10_Zb_2_1_R}{C_01_C_01_R}{C_10_C_11_R}{}{}
        \transitionSector[angle left=5, angle right=5]{Yb_2_00_Zb_2_1_R}{C_00_C_10_R}{C_00_C_11_R}{}{}
        \transitionSector[angle left=5, angle right=5]{Yb_2_10_Zb_2_0_R}{C_10_C_00_R}{C_10_C_01_R}{}{}
        \transitionSector[angle left=5, angle right=5]{Yb_2_11_Zb_2_1_R}{C_11_C_10_R}{C_01_C_01_R}{}{}
        \transitionSector[angle left=5, angle right=5]{Yb_2_00_Zb_2_0_R}{C_01_C_01_R}{C_00_C_01_R}{}{}
        \transitionSector[angle left=5, angle right=5]{Yb_2_10_Zb_2_1_L}{C_10_C_10_L}{C_10_C_11_L}{}{}
        \transitionSector[angle left=5, angle right=5]{Yb_2_01_Zb_2_0_L}{C_01_C_00_L}{C_01_C_01_L}{}{}
        \transitionSector[angle left=5, angle right=5]{Yb_2_10_Zb_2_0_L}{C_10_C_00_L}{C_01_C_10_L}{}{}
        \transitionSector[angle left=5, angle right=5]{Yb_2_01_Zb_2_1_R}{C_01_C_10_R}{C_01_C_11_R}{}{}
        \transitionSector[angle left=5, angle right=5]{Yb_2_11_Zb_2_1_L}{C_10_C_11_L}{C_11_C_11_L}{}{}
        \transitionSector[angle left=5, angle right=5]{Yb_2_11_Zb_2_0_L}{C_00_C_11_L}{C_01_C_11_L}{}{}
        \transitionSector[angle left=5, angle right=5]{Yb_2_00_Zb_2_0_L}{C_00_C_00_L}{C_01_C_00_L}{}{}
        \transitionSector[angle left=5, angle right=5]{Yb_2_11_Zb_2_0_R}{C_11_C_00_R}{C_11_C_01_R}{}{}
        \transitionSector[angle left=5, angle right=5]{Yb_2_01_Zb_2_0_R}{C_01_C_00_R}{C_01_C_01_R}{}{}
        \transitionSector[angle left=5, angle right=5]{Yb_2_01_Zb_2_1_L}{C_01_C_10_L}{C_01_C_11_L}{}{}
        \transitionSector[angle left=0, angle right=0, shift={0, 0.9cm}]{C_01_C_00_R}{C_1_C_0_R}{O_1_O_1_L}{}{}
        \transitionSector[angle left=0, angle right=0, shift={0, 0.9cm}]{C_01_C_00_L}{C_1_C_0_L}{O_1_O_1_L}{}{}
        \transitionSector[angle left=0, angle right=0, shift={0, 0.9cm}]{C_11_C_00_R}{O_1_C_0_R}{C_1_O_1_L}{}{}
        \transitionSector[angle left=0, angle right=0, shift={0, 0.9cm}]{C_00_C_00_L}{C_0_C_0_L}{O_1_O_1_L}{}{}
        \transitionSector[angle left=0, angle right=0, shift={0, 0.9cm}]{C_00_C_11_R}{O_1_C_0_L}{O_1_C_1_R}{}{}
        \transitionSector[angle left=0, angle right=0, shift={0, 0.9cm}]{C_01_C_01_R}{O_1_O_1_L}{O_1_O_1_L}{}{}
        \transitionSector[angle left=0, angle right=0, shift={0, 0.9cm}]{C_10_C_00_R}{O_1_C_0_R}{O_1_C_0_L}{}{}
        \transitionSector[angle left=0, angle right=0, shift={0, 0.9cm}]{C_10_C_11_L}{O_1_O_1_L}{C_1_C_0_L}{}{}
        \transitionSector[angle left=0, angle right=0, shift={0, 0.9cm}]{C_10_C_10_L}{O_1_O_1_L}{C_0_C_0_L}{}{}
        \transitionSector[angle left=0, angle right=0, shift={0, 0.9cm}]{C_11_C_11_L}{O_1_O_1_L}{C_1_C_1_L}{}{}
        \transitionSector[angle left=0, angle right=0, shift={0, 0.9cm}]{C_00_C_11_L}{O_1_C_0_L}{C_1_O_1_L}{}{}
        \transitionSector[angle left=0, angle right=0, shift={0, 0.9cm}]{C_00_C_01_R}{C_0_C_1_R}{O_1_O_1_L}{}{}
        \transitionSector[angle left=0, angle right=0, shift={0, 0.9cm}]{C_10_C_11_R}{O_1_O_1_L}{C_0_C_1_R}{}{}
        \transitionSector[angle left=0, angle right=0, shift={0, 0.9cm}]{C_01_C_10_L}{C_1_O_1_L}{O_1_C_0_L}{}{}
        \transitionSector[angle left=0, angle right=0, shift={0, 0.9cm}]{C_00_C_10_R}{O_1_C_0_L}{O_1_C_0_R}{}{}
        \transitionSector[angle left=0, angle right=0, shift={0, 0.9cm}]{C_11_C_01_R}{O_1_C_1_R}{C_1_O_1_L}{}{}
        \transitionSector[angle left=0, angle right=0, shift={0, 0.9cm}]{C_01_C_01_L}{C_1_C_1_L}{O_1_O_1_L}{}{}
        \transitionSector[angle left=0, angle right=0, shift={0, 0.9cm}]{C_01_C_11_R}{C_1_O_1_L}{O_1_C_1_R}{}{}
        \transitionSector[angle left=0, angle right=0, shift={0, 0.9cm}]{C_11_C_10_R}{O_1_O_1_L}{C_1_C_0_R}{}{}
        \transitionSector[angle left=0, angle right=0, shift={0, 0.9cm}]{C_10_C_01_R}{O_1_C_1_R}{O_1_C_0_L}{}{}
        \transitionSector[angle left=0, angle right=0, shift={0, 0.9cm}]{C_01_C_11_L}{C_1_O_1_L}{C_1_O_1_L}{}{}
        \transitionSector[angle left=0, angle right=0, shift={0, 0.9cm}]{C_01_C_10_R}{C_1_O_1_L}{O_1_C_0_R}{}{}
        \transitionSector[angle left=0, angle right=0, shift={0, 0.9cm}]{C_10_C_00_L}{O_1_C_0_L}{O_1_C_0_L}{}{}
        \transitionSector[angle left=5, angle right=5]{O_1_O_1_L}{O_0_O_0_L}{O_0_O_0_L}{}{}
        \transitionSector[angle left=5, angle right=5]{O_1_C_1_R}{O_0_O_0_L}{O_0_C_R}{}{}
        \transitionSector[angle left=5, angle right=5]{C_1_C_0_L}{C_O_0_L}{C_O_0_L}{}{}
        \transitionSector[angle left=5, angle right=5]{C_1_O_1_L}{O_0_O_0_L}{C_O_0_L}{}{}
        \transitionSector[angle left=5, angle right=5]{C_0_C_1_R}{C_O_0_L}{O_0_C_R}{}{}
        \transitionSector[angle left=5, angle right=5]{C_0_C_0_L}{C_C_L}{O_0_O_0_L}{}{}
        \transitionSector[angle left=5, angle right=5]{O_1_C_0_L}{C_O_0_L}{O_0_O_0_L}{}{}
        \transitionSector[angle left=5, angle right=5]{C_1_C_1_L}{O_0_O_0_L}{C_C_L}{}{}
        \transitionSector[angle left=5, angle right=5]{C_1_C_0_R}{O_0_C_R}{C_O_0_L}{}{}
        \transitionSector[angle left=5, angle right=5]{O_1_C_0_R}{O_0_C_R}{O_0_O_0_L}{}{}

        \draw[thick] (O_0_C_R) -- ++(0,1cm) node[below, font=\tiny] {$-\frac{1}{8}$};
        \draw[thick] (C_C_L) -- ++(0,1cm) node[below, font=\tiny] {$\frac{1}{4}$};
        \draw[thick] (C_O_0_L) -- ++(0,1cm) node[below, font=\tiny] {$\frac{1}{8}$};
        \draw[thick] (O_0_O_0_L) -- ++(0,1cm) node[below, font=\tiny] {$0$};
    \end{tikzpicture}
    \caption{The LSTA representing $T_g^{\pi_{\mathrm{sel}}}$ after applying $H_t$ to the selector input in \cref{fig:selector-input}.
    Equal residuals at each level have been merged.
    Only the initial state name and amplitude labels are shown: every choice set is $\{0\}$, and successive internal levels have labels $t,y_1,z_1,y_2,z_2,w_1,w_2$.}
    \label{fig:selector-hadamard}
\end{figure}
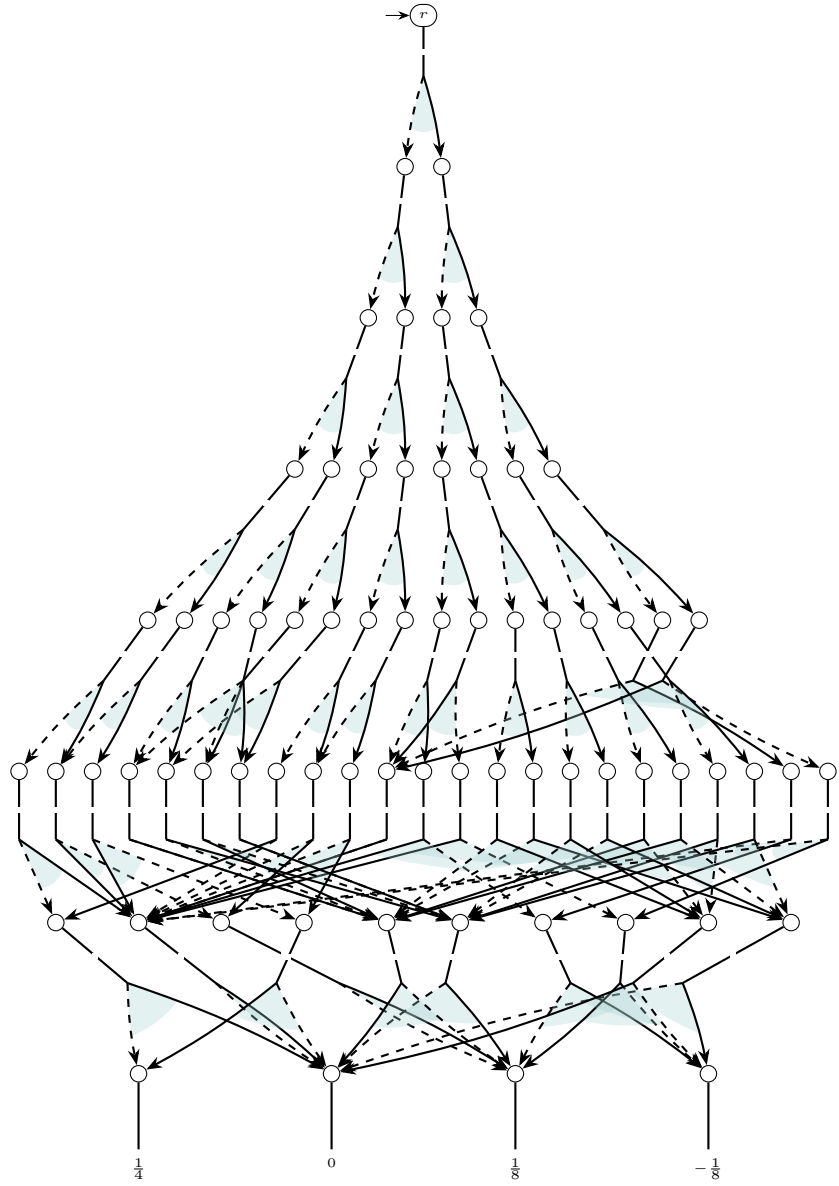

\end{document}